\documentclass[11pt]{article}
\usepackage{jheppub}
\usepackage{amsmath,amssymb,amsfonts,graphicx,slashed,color}

\usepackage{color,graphicx}

\newcommand{\ep}{\varepsilon}
\newcommand{\fH}{H^f} %Full modular Hamiltonian

\newcommand{\pa}{\partial}

\newcommand{\beq}{\begin{equation}}
\newcommand{\eeq}{\end{equation}}
\newcommand{\beqn}{\begin{eqnarray}}
\newcommand{\eeqn}{\end{eqnarray}}

\newcommand{\bs}{\boldsymbol}
\newcommand{\bbdy}{B}
\newcommand{\cS}{\mathcal{S}}
\newcommand{\bdx}{\mathbf{x}}

\newcommand{\cO}{\mathcal{O}}

\newcommand{\cL}{\mathcal{L}}
\title{The Holographic Shape of Entanglement and Einstein's Equations}
                                           \author[\,a]{Aitor Lewkowycz}
                                           \author[\,b]{,\;Onkar Parrikar}
                                           
                                           \affiliation[a]{Stanford Institute for Theoretical Physics, Deptartment of Physics\\
                                           Stanford University, Stanford, CA 94305, USA.}
                                            \affiliation[b]{David Rittenhouse Laboratory, University of Pennsylvania\\
                                            209 S. 33rd Street, Philadelphia, PA 19104, USA.}
                                           
                          \emailAdd{lewkow@stanford.edu, parrikar@sas.upenn.edu}
                                           
 \abstract{We study shape-deformations of the entanglement entropy and the modular Hamiltonian for an arbitrary subregion and state (with a smooth dual geometry) in a holographic conformal field theory. More precisely, we study a double-deformation comprising of a shape deformation together with a state deformation, where the latter corresponds to a small change in the bulk geometry. Using a purely gravitational identity from the Hollands-Iyer-Wald formalism together with the assumption of equality between bulk and boundary modular flows for the original, undeformed state and subregion, we rewrite a purely CFT expression for this double deformation of the entropy in terms of bulk gravitational variables and show that it precisely agrees with the Ryu-Takayanagi formula including quantum corrections. As a corollary, this gives a novel, CFT derivation of the JLMS formula for arbitrary subregions in the vacuum, without using the replica trick. Finally, we use our results to give an argument that if a general, asymptotically AdS spacetime satisfies the Ryu-Takayanagi formula for arbitrary subregions, then it must necessarily satisfy the non-linear Einstein equation.}


\begin{document}
                                           
\maketitle

\parskip=10pt
\section{Introduction} 
A lot of progress has been made in recent years in understanding the gravity dual of entanglement entropy in holographic conformal field theories
 \cite{Ryu:2006bv,Hubeny:2007xt, Casini:2011kv, Lewkowycz:2013nqa,Dong:2013qoa,Camps:2013zua,Dong:2016hjy}. So far, much of this work has focussed on using the replica trick (except for \cite{Casini:2011kv}, which leverages the symmetries of ball-shaped regions in the vacuum), which in quantum field theories requires putting the theory on a different background with a conical deficit at the entangling surface, together with other subtle operations such as analytic continuation in the replica index. It is desirable to gain further understanding of holographic entanglement entropy using more direct techniques, given that it should be computable directly within the original Hilbert space. There are several motivations for this -- firstly, it could potentially provide a clearer understanding of the meaning of subregions in quantum gravity (in AdS) and could provide further insight into the microscopic origin of the Bekenstein-Hawking entropy, perhaps in terms of counting of edge modes \cite{Donnelly:2016auv,Speranza:2017gxd}. Another motivation would be to give a more direct derivation of the Ryu-Takayanagi (RT) formula, without using the replica trick. Finally, there has been much work in recent years suggesting a deep connection between the emergence of spacetime geometry and entanglement in the AdS/CFT correspondence \cite{VanRaamsdonk:2010pw, Lashkari:2013koa, Faulkner:2013ica, Nozaki:2013vta}. For instance, it was shown in these papers that any asymptotically AdS spacetime which computes the entanglement entropies for ball-shaped regions in the CFT using the Ryu-Takayanagi formula for up to first order state deformations around the vacuum, necessarily satisfies the linearized Einstein equation around AdS. It is likely that understanding this connection further will involve essentially new techniques. 
 %Furthermore, while in the replica trick, one defines the entropy by going to an auxiliary spacetime , one expects that there is some derivation of the RT formula in terms of properties of the state itself. 

%In this paper, we are going to try to make some progress in the previous direction by focusing on small variations around arbitrary backgrounds.
One approach along these lines is entanglement (or modular) perturbation theory, where one studies the entanglement entropy (or correlation functions of the modular Hamiltonian) perturbatively around a background state, for small deformations in the state or shape of the subregion. This approach was first explored in \cite{Rosenhaus:2014woa}, and later improved upon in \cite{Faulkner:2014jva}. Since then, there have been many advances, especially when the perturbations are shape deformations \cite{Allais:2014ata,Rosenhaus:2014zza,Lewkowycz:2014jia,Mezei:2014zla,Bianchi:2015liz, Faulkner:2015csl,Faulkner:2016mzt,Dong:2016wcf, Balakrishnan:2016ttg, Bianchi:2016xvf, Casini:2017roe, Lashkari:2017rcl, Sarosi:2017rsq}. In addition to being computationally useful, entanglement perturbation theory has found several interesting applications. For instance, \cite{Faulkner:2016mzt} derived the averaged null energy condition (ANEC) in Minkowski spacetime from the monotonicity of relative entropy together with entanglement perturbation theory for shape deformations (see also \cite{Hartman:2016lgu} for another proof of the ANEC using OPE techniques). Entanglement perturbation theory was also used \cite{Faulkner:2017tkh, Haehl:2017sot} to derive the gravitational equations of motion from entanglement to second order around AdS. Importantly, much of this work so far has focussed on special symmetric situations, such as for instance, deformations around ball-shaped regions in the CFT vacuum, which in the holographic context only probes small deformations around AdS spacetime. Further progress necessitates moving away from such special cases.  

In this paper, we will take the first steps in this direction by studying the shape deformations of entanglement entropy for a general region $R$ and a general state $\psi$ (with a smooth AAdS dual geometry $g$) in a holographic conformal field theory with Einstein gravity dual (although our techniques can also be applied to higher-curvature theories). More precisely, we will be interested in a double-deformation $\delta \delta_VS$ of the entanglement entropy, where $\delta_V$ denotes a shape deformation of the subregion, while $\delta$ denotes a state deformation around the reference state.  From the CFT point of view, we have the following boundary expression for this double deformation of the entropy:
\begin{equation}
\delta \delta_V S = \lim_{\bbdy \to 0} \int_{-\infty}^{\infty} \frac{ds}{4\sinh^2(\frac{s+i\epsilon}{2})} \oint_{\partial R_\bbdy} V^{\mu} n^{\nu} \delta \left \langle \rho_R^{-is/2\pi} \,:T_{\mu\nu}: \,\rho_R^{is/2\pi} \right \rangle \label{d2Sintro}
\end{equation}
where $\partial R_B$ is a small (Euclidean) tube of radius $B$ which surrounds the entangling surface, and $V$ is the vector field parametrizing the shape deformation.  This formula essentially follows from the setup in \cite{Faulkner:2016mzt} and will be explained in more detail in section \ref{sec2}, but at this point we would like to highlight a few of its salient properties. Firstly, it contains the evolution operator $\rho_R^{is/2\pi}$ involving the density matrix of $\psi$ reduced over $R$, which generates what is commonly called \emph{modular flow}; this is crucial for the right hand side to have a non-trivial limit as $B \to 0$. Secondly,  it depends on the integral of the stress tensor on a co-dimension one surface $\partial R_B$ (which naively becomes co-dimension $2$ as $B\rightarrow 0$), which greatly facilitates rewriting it in terms of bulk gravitational variables. Finally, equation \eqref{d2Sintro} provides a purely field-theoretic \emph{constraint} on a particular deformation of the entropy (on the left hand side) in terms of the stress tensor expectation value (on the right hand side), which we will see has an interesting manifestation in the bulk. 

Indeed, for holographic theories dual to Einstein gravity, we expect this double-deformation of the entanglement entropy to be computed by the change in the area of the bulk extremal surface:
\begin{equation}
\delta \delta_V S =\frac{\delta \delta_V A_{ext} }{4 G_N } \label{d2Aintro}.
\end{equation}
One of the goals of this paper is to derive \eqref{d2Aintro} from \eqref{d2Sintro}. In fact, we will also derive the quantum corrections to the above formula coming from the bulk entanglement entropy \cite{Faulkner:2013ana}, but these have been omitted here for simplicity. Our derivation below will use two main ingredients: $(i)$ a purely gravitational identity from the Hollands-Iyer-Wald formalism \cite{Iyer:1995kg, Hollands:2012sf} (which can be thought of as the gravitational equivalent of Gauss' law), which, upon using the extrapolate dictionary, allows us to rewrite the right hand side of equation \eqref{d2Sintro} in terms of bulk geometric quantities, and $(ii)$ we will assume that for the original (i.e., undeformed) reference state $\psi$ and subregion $R$, boundary modular flow is equivalent to bulk modular flow.\footnote{This is very natural as it essentially amounts to a matching of bulk and boundary symmetries in the background.} These two ingredients, along with the bulk equations of motion (linearized around the background geometry) will directly lead to a derivation of equation \eqref{d2Aintro}. Since $R$ and $\delta \psi$ were arbitrary, this equation can then be bootstrapped to large shape-deformations. For instance, as a corollary to equation \eqref{d2Aintro}, we can give a novel, perturbative, CFT derivation of the JLMS formula \cite{Jafferis:2015del} for an arbitrary subregion (with the topology of a ball) in the vacuum state, without using the replica trick. 

In fact, we can also go in the opposite direction -- assuming that the bulk geometry satisfies equation \eqref{d2Aintro}, we will be able to give an argument %(albeit not a proof) 
that it must also satisfy the linearized Einstein equation around the background $g$. Since $g$ can be taken to be an arbitrary AAdS solution to the Einstein equation, this then implies that any asymptotically AdS spacetime which satisfies the Ryu-Takayanagi formula for arbitrary subregions, necessarily satisfies the non-linear Einstein equation! Indeed, as we will see in more detail later, our derivation will involve an interesting interplay between the equations of motion, extremality and modular flow.
%\AL{Added this, it is a little redundant but it might be helpful for the reader to find the assumptions and results directly there}
\subsubsection*{Summary of results}
For clarity, we will presently state our results and the assumptions we will use to derive them. We will work with a general subregion and state (with a smooth bulk dual) in a holographic conformal field theory. We will throughout assume: (i) the extrapolate dictionary near the asymptotic boundary, and (ii) the equality between the \emph{background} (i.e., corresponding to the undeformed state and subregion) bulk and boundary modular flows. With these assumptions, we can re-write the purely CFT expression \eqref{d2Sintro} in terms of bulk gravitational variables, as discussed in section \ref{sec3}. This general bulk expression then has the following properties:
\begin{itemize}
\item If we assume the linearized bulk equations of motion around $g$, then we obtain a derivation of the RT (JLMS) formula (for small state variations), upon requiring the perturbed surface to be extremal. For the special case of the vacuum, since the equality between bulk and boundary modular flows for ball-shaped regions was derived in \cite{Casini:2011kv}, this is a first-principles derivation of the perturbative RT formula for an arbitrary region (with the topology of a ball) in the vacuum. For more general states, our resuts imply the perturbative RT formula for an arbitrary region assuming the equality between modular flows for a reference region \footnote{As we will explain later, the equality between modular flows is a statement about the locality of the holographic mapping and it doesn't require any reference to the area of the surface.}.   
\item If we assume that the equality between bulk and boundary flows continues to hold even for the shape-deformed subregion (together with the bulk equations of motion), then we obtain a derivation of the extremality condition. 
\item On the other hand, if we do not assume the equations of motion and instead assume the RT formula (or equivalently JLMS), then the bulk expression for \eqref{d2Sintro} is only compatible with JLMS if the linearized equations of motion around $g$ are satisfied along the RT surface. This gives a derivation of the linearized equations of motion around an \emph{arbitrary} AAdS background. 
\end{itemize}

The rest of the paper is organized as follows. In section \ref{sec2} we will introduce the necessary background material and set up notation. In section \ref{sec3}, we will present our results about the gravitational dual of \eqref{d2Sintro}. We will then use it to prove equation \eqref{d2Aintro}, and also explain the various implications for JLMS, extremality, bulk equations of motion etc. We finish with some closing comments in section \ref{sec4}. In order to avoid cluttering the main body of the paper, we defer most of the detailed calculations to the various appendices. 

\section{Preliminaries} \label{sec2}
In this section, we will review some of the prerequisite background material and set up notation which will be used through the rest of the paper.
\subsection{Perturbative approach to entanglement}
Let us begin by considering a general state $\psi$ in a relativistic quantum field theory, and let $R$ be a general subregion on a Cauchy surface $\Sigma$. Assuming that the Hilbert space of the theory on $\Sigma$ factorizes, the reduced density matrix corresponding to $\psi$ on the subregion $R$ is given by 
\beq
\rho_{\psi, R} = \mathrm{Tr}_{R^c} |\psi\rangle \langle \psi |.
\eeq
The entanglement entropy is then defined as the von Neumann entropy of this density matrix:
\beq \label{EE}
S(\rho_{\psi, R} ) = - \mathrm{Tr}_{R}\left(\rho_{\psi, R}\,  \ln\,\rho_{\psi, R}\right).
\eeq
It is also convenient to define the \emph{modular Hamiltonian} as
\beq \label{MH}
H_{\psi,R} \equiv -\log \rho_{\psi,R},
\eeq
in terms of which the reduced density matrix takes the thermal form $\rho_{\psi,R} = e^{-H_{\psi,R}}$. In special circumstances, the modular Hamiltonian is local, i.e. its action on local operators is local and geometric. For instance, if $R$ is a half-space in the vacuum of a relativistic quantum field theory, then
\beq
i\left[ H_{\psi, R} , \cO_{\mu_1,\cdots,\mu_n}(x) \right ] = \xi^{\nu}(x)\pa_{\nu} \cO_{\mu_1,\cdots,\mu_n}(x)+ w_\cO\, \cO_{\mu_1,\cdots,\mu_n}(x),
\eeq
where $\xi^{\mu} $ is the vector field corresponding to boosts around the entanglement cut, and $w_\cO$ is the \emph{modular weight} of the operator $\cO$. Consequently, in this case $\rho_{\psi, R}^{is/2 \pi} = e^{-i \frac{s}{2\pi} H_{\psi,R}}$ generates a geometric \emph{modular flow}. Note however, that this is only true in very special situations, and for the case of general $\psi$ and $R$ which is of interest in the present paper, the modular Hamiltonian is not local. Nevertheless, modular evolution maps the algebra of operators inside $R$ into itself. In order to avoid cluttering notation, we will henceforth drop the subscript $\psi$ on the density matrix and simply write $\rho_R$.

Entanglement perturbation theory is a useful tool for computing the entanglement entropy perturbatively for small deformations around a reference state/subregion \cite{Rosenhaus:2014woa}. For example, given a small deformation of the reference state $\delta |\psi \rangle$ (and the corresponding change $\delta \rho_{R}$ in the reduced density matrix), the first order change in the entanglement entropy is given by
\beq
\delta S = \mathrm{Tr}_{R} \left(\delta \rho_{ R} \,H_{ R} \right) \equiv  \langle H_{R} \rangle_{\delta \psi}.
\eeq 
This is known as the \emph{first law} of entanglement entropy, and has found many interesting applications \cite{Blanco:2013joa, Faulkner:2013ica}. Consider now, a second order variation of the entropy $\delta_1\delta_2 S$, where $\delta_1$ and $\delta_2$ denote the two (generically different) variations. From the definitions \eqref{EE} and \eqref{MH}, we obtain\footnote{Note that we want the state to be normalized, which is equivalent to $\text{tr} (\rho \delta \ln \rho) =\text{tr} \delta \rho=0$, for any $\delta$, so there is no such contribution in the expansion. In equation \eqref{deltaS}, we have used this after the first variation.} :
%\footnote{Note that because of the normalization of the density matrix ${\delta \rho}=0$, the terms without $\delta \rho$ are zero. [MAKE SURE]} :
\begin{equation}
\delta_1 \delta_2 S = \langle H_{R} \rangle_{\delta_1 \delta_2 \psi}+ \langle \delta_1  H_{R} \rangle_{\delta_2 \psi}. \label{deltaS}
\end{equation}
The first term on the right hand side is similar to the contribution we found at first order. The latter  containing $\delta_{1} H$ is more subtle and interesting. In fact, this term is equal to the \emph{relative entropy} between the two density matrices $\tilde{\rho}_R=\rho_R+\delta_1 \rho_R+\delta_2 \rho_R$ and $\rho_R$:
\begin{equation}
S_{rel}(\tilde{\rho}_R|\rho_R)\equiv \langle H_{R} \rangle_{\delta_1 \delta_2 \psi}-\delta_1 \delta_2 S(\rho_{R})=\langle \delta_1  H_{R} \rangle_{\delta_2 \psi},
\end{equation}
and as such is symmetric between $\delta_1$ and $\delta_2$, despite appearances. Relative entropy in quantum field theory is generally expected to be free of the ultraviolet (UV) divergences typically found in the entanglement entropy. Because of this fact, we expect terms of the form $\langle \delta_1  H_{R} \rangle_{\delta_2 \psi}$ to be UV finite, which is an extra motivation to study them. Such terms will be our central focus in the present paper.

In order to proceed, we need to compute $\delta H_{R}$ in terms of $\delta \rho_{R}$. Since $H_{R} = - \ln\,\rho_{R}$, we can use the Baker-Campbell-Hausdorff formula to compute $\delta H_{R}$ (see \cite{Faulkner:2016mzt}):
\begin{equation}
\delta H_{R} = \int_{-\infty}^{\infty} ds \frac{1}{4 \sinh^2\left(\frac{s+i \epsilon}{2} \right)}  \rho_{R}^{-i s/2\pi}\,\rho_R^{-1}\delta \rho_{R} \,\rho_{R}^{i s/2\pi}. \label{deltaK}
\end{equation}
The parameter $s$ denotes ``modular {flow}'' -- when the modular hamiltonian is local, for instance in the case of a half-space in the vacuum, $s$ parametrizes Rindler-time evolution. More generally, this internal evolution is very non-local and hard to compute explicitly, but it is nonetheless a useful construct \cite{Faulkner:2017vdd}, among other reasons because it is a well defined operation even in the continuum limit. This can be seen by considering the full modular hamiltonian: %\ack{Replace $\bar{H}$ with $H^{f}$}
\begin{equation}
\fH_{\partial R} \equiv H_{R}-H_{R^c}.
\end{equation}
This operator is well defined: it does not present the ambiguities near $\partial R$ which the entropy has and the modular flow is more generally defined as being generated by the full modular Hamiltonian:
\begin{equation}
O(s) \equiv e^{i \frac{s}{2\pi} \fH_{\partial R}} \,O\, e^{-i \frac{s}{2\pi} \fH_{\partial R} }.
\end{equation}

With equation (\ref{deltaK}), one can in principle re-write and compute (\ref{deltaS}) in terms of correlation functions in the reference state $\psi$, provided the modular evolution in \eqref{deltaK} can be carried out explicitly, as was done, for instance, in \cite{Faulkner:2014jva, Faulkner:2015csl}. However, since we are interested in the general situation where modular flow is not local, we will not have this luxury. In order to make progress in this situation, we will consider the particular case where one of the variations -- which we will henceforth call $\delta_V$ -- is a deformation of the shape of the entangling surface. The other variation -- which we will simply call $\delta$ -- will be taken to be a state deformation, which in the holographic context corresponds to a small deformation of the bulk geometry. 

\subsection{Shape deformations}
In this section we review several details about the shape dependence of entanglement entropy and modular Hamiltonians, first for a general CFT and then from a gravitational perspective for holographic CFT's with Einstein gravity duals.

\subsection*{CFT perspective}
As was discussed, for example in \cite{,Allais:2014ata,Faulkner:2016mzt}, we can think of a shape deformation in terms of shifting the background metric by a pure diffeomorphism; we will now briefly review this argument. We can construct the reference state $\psi$ by performing the Euclidean path-integral on the lower half-space in Euclidean signature ($x^0  < 0$, where $x^0$ is Euclidean time), with some (not necessarily local\footnote{For instance, we could choose $\Psi = \sum_\alpha c_{\alpha} O_{\alpha}(x^0_{\alpha},\bdx_{\alpha})$, where $x^0_{\alpha} < -t,\;\forall\;\alpha$,  or even consider states prepared by turning on Euclidean sources for certain operators. }) operator insertion $\Psi$ away from $x^0 = 0$:
\beq
\langle \varphi (\bdx) | \psi \rangle = \int^{\phi(x^0 = 0,\bdx) = \varphi(\bdx)} [D\phi] e^{-S_{CFT}[\phi] } \;\Psi,
\eeq 
where $\phi$ collectively denote the elementary fields which are integrated over in the path-integral, and $\varphi$ denote the boundary conditions on the Cauchy surface $x^0=0$ (with $\bdx$ being spatial coordinates on this surface). Note that for states in holographic CFTs dual to smooth bulk geometries, it might be more natural to consider coherent states, where we turn on (not necessarily small) sources for single-trace, primary operators in the Euclidean path-integral.  As long as the sources only have support far from $x^0 = 0$, our arguments carry over in this case as well. The reduced density matrix $\rho_R$ can then be obtained by gluing this path-integral with its image under time-reversal along the complement region $R^c$:
 \beq\label{dm}
\langle \varphi_R^-(\bdx) | \rho_R| \varphi_R^+(\bdx) \rangle = \frac{1}{Z}\int^{\phi(x^0 = 0^+,\bdx) = \varphi_R^+(\bdx)}_{\phi(x^0 = 0^-,\bdx) = \varphi_R^-(\bdx)} [D\phi] e^{-S_{CFT}[\phi] }\;\Psi^{\dagger}\Psi,
\eeq 
where now $\varphi_R^{\pm}$  are boundary conditions over and under the cut along $R$, and $Z$ normalizes the density matrix. Let $V$ denote the vector field which parametrizes the deformation of the entangling surface; $V$ is a priori defined on $\pa R$, but we can pick some smooth extension to a neighborhood of $\pa R$. Now, by performing a diffeomorphism $x^{\mu} \to x^{\mu} - V^{\mu}$, we can map the deformed entangling surface to the original one. Of course, the diffeomorphism acts non-trivially on the background metric:
\beq \label{dmet}
\delta_V g_{\mu\nu} = 2 \pa_{(\mu}V_{\nu)},
\eeq 
and so in this way we can trade the shape-deformation of the entangling surface with a metric deformation:
\begin{equation}
\rho_{R+\delta_V R,\eta}=U^{\dagger} \rho_{R,\eta+\delta_V g} U
\end{equation}
where $U$ is a unitary map from the Hilbert space of the subregion $R+\delta_V R$ to the hilbert space in the subregion $R$. Consequently, the variation of the modular hamiltonian has two contributions: (i) from the variation of $-\ln \,\rho_{R,\eta+\delta_Vg}$ coming from the change in the metric $\delta_Vg$, and (ii) from the unitary transformation $U$ (which also depends on $V$). 

\begin{figure}[t]
\centering
\includegraphics[height=5cm]{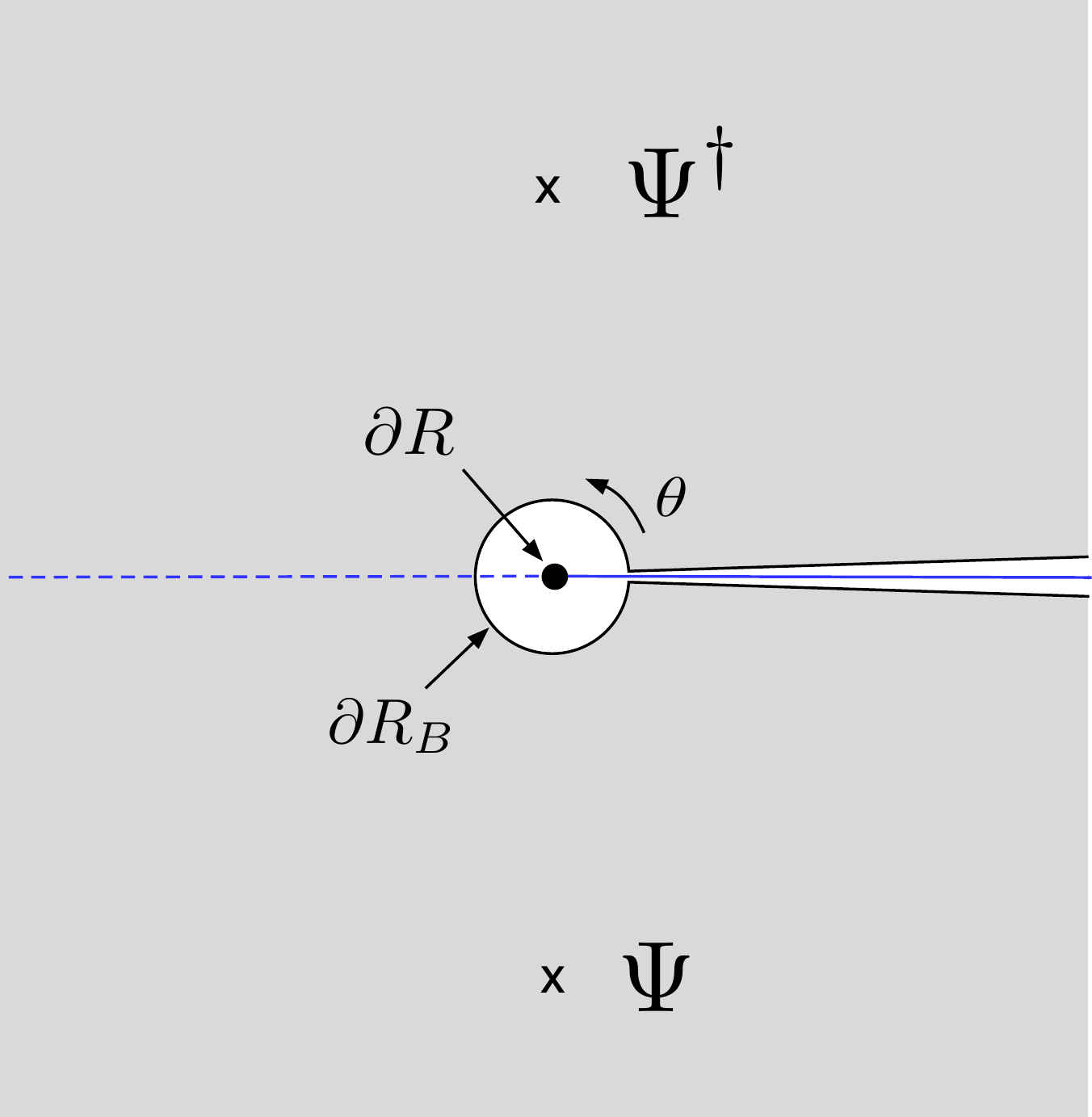}
\caption{\small{An illustration of the Euclidean path integral representation of the reduced density matrix. The solid blue line denotes the region $R$, the dashed blue line is $R^c$, and the solid black dot is $\pa R$. Also shown are the cylindrical tube surrounding the entangling surface $\pa R_B$, and the cut at $\theta = 0$. } \label{fig1}}
\end{figure}

For contributions of type (i), the variation of the density matrix $\delta_g\rho_R$ with respect to the Euclidean metric deformation \eqref{dmet} can be read off from equation \eqref{dm}:
 \beq\label{dgdm}
\langle \varphi_R^- | \delta_g \rho_R| \varphi_R^+ \rangle = \frac{1}{Z}\int^{\phi(x^0 = 0^+) = \varphi_R^+}_{\phi(x^0 = 0^-) = \varphi_R^-} [D\phi] e^{-S_{CFT}}\frac{1}{2}\int d^dx\; \delta_Vg^{\mu\nu}(x):T_{\mu\nu}:(x)\;\Psi^{\dagger}\Psi,
\eeq 
where we have defined
\begin{equation}
:O:\;\equiv\; O-\langle O \rangle_{\psi}.\;\;\; %O(s)\equiv \rho_{R}^{-i s/2 \pi} O \rho_{R}^{i s/2\pi}
\end{equation}
The upshot of this discussion is that inside the path-integral, we can make the following replacement for $\rho_R^{-1}\delta_g \rho_R$: 
\begin{equation}\label{dm2}
\rho_R^{-1}\delta_g \rho_R= \int d^d x \,\partial^{\mu} V^{\nu}(x) \,:T_{\mu \nu}:(x),
\end{equation}
%or equivalently
%\beq
%\mathrm{Tr}_R\left(\delta_g\rho \cdots\right)= \int d^d x \,\partial^{\mu} V^{\nu}(x) \left\langle:T_{\mu \nu}:(x) \cdots \right\rangle_{\psi}.
%\eeq
%where $\cdots$ are some operator insertions along $R$. 
where note that $T_{\mu\nu}$ is inserted in Euclidean time, and hence should be interpreted as a non-local, Heisenberg operator in terms of the operator-algebra on $R$. Now we can integrate by parts in $x$. In \cite{Faulkner:2016mzt}, it was shown that this gives two contributions: one from a $(d-1$) dimensional tube $\pa R_{\bbdy}$ of radius $\bbdy$ which surrounds the entangling surface, and another contribution from the cut along the region $R$ on the $x^0=0$ slice (see figure \ref{fig1}). %\ack{I think we should distinguish between the boundary radius and the bulk radius; they are related by a factor of $z$ in changing to FG gauge.} \AL{Probably prefer to not have a subindex, we could use $B$, which is ugly yet consistent with the notation, just change "bbdy"}
 In fact, this latter contribution coming from the cut exactly cancels the contribution of type (ii) above, coming from the unitary transformations $U$.  Therefore, the final change in the modular hamiltonian is given by: 
\begin{equation}
 \langle \delta_V H_{R} \rangle_{\delta \psi}  = \lim_{\bbdy \to 0} \int_{-\infty}^{\infty} \frac{ds}{4\sinh^2(\frac{s+i\epsilon}{2})} \oint_{\partial R_\bbdy} V^{\mu} n^{\nu} \langle \rho_R^{-is/2\pi} \,:T_{\mu\nu}:(\bbdy,\theta, y) \,\rho_R^{is/2\pi} \rangle_{\delta \psi}, \label{dXK}
\end{equation}
where $y$ denotes the coordinates along the entangling surface, $\partial R$, and $\theta$ is the angular coordinate around the tube (see figure \ref{fig1} for an illustration).  Note that $\partial R_\bbdy$ has a cut along $\theta = 0$ (i.e. $\varepsilon \leq \theta \leq 2\pi - \varepsilon$) and $\bbdy$ should be regarded as much smaller than any of the curvature scales in the CFT, because, we are interested in the limit $\bbdy \to 0$. Note that equation \eqref{dXK} is true for general subregions and does not require a $U(1)$ rotation symmetry around the entanglement cut (although reference \cite{Faulkner:2016mzt}, where it was derived, focused on situations where such a symmetry is present\footnote{We are using a definition for $\delta_V H_R$ where this operators live in the hilbert space of the subregion $R+\delta_V R$, which is more convenient for our purposes. In particular, $\langle \delta_V H \rangle_{\psi} \not =0$ and this term is responsible for the change in the entropy, via the contact term described in the next paragraph. This definition of $\delta_V H_R$ is directly connected with the area operator, since it is a state independent operator, but its shape variation is non-zero (it changes the boundary conditions).}). 

Additionally, there is an extra contact term at the entangling surface that should be added to the $\delta_V H_R$ operator, which is discussed in Appendix A of \cite{Faulkner:2016mzt}. This term is important and it is the only contribution to the change in the entropy due to a shape deformation \cite{Allais:2014ata}. However, since our double-deformation is unambiguous, it should not depend on this contact term; in other words, we expect this contact term to be state-independent.\footnote{This contact term is related with the $T_{\mu\nu} H_R$ OPE. In holographic theories, we can rewrite this term in terms of bulk gravitational variables (by using techniques to be discussed later), and the only way it can survive is if we encounter bulk UV divergences to counter the suppression in the $B\to 0$ limit. However, since $H_R$ is a UV finite operator from the bulk point of view, we expect this not to happen. From the point of view of the change in the area, we can understand this state independence as the fact that there is a boundary term in the asymptotic $AdS$ boundary which is non-zero in the background state but vanishes upon doing a state deformation, see appendix B for more details. } Further, this contact term should cancel out in the case of the full modular Hamiltonian $\delta_V H^{f}_{\pa R}$, while the right hand side of \eqref{dXK} will survive even in the full modular Hamiltonian.    

Coming back to \eqref{dXK}, naively it might seem that this term vanishes in the $\bbdy \to 0$ limit, but in fact the $s$-integral gives an enhancement from the integration regions $s \sim \pm \ln \bbdy$, thus leading to a finite result \cite{Faulkner:2016mzt}.\footnote{For instance in the case of a half-space in the vaccum of a relativistic quantum field theory, equation \eqref{dXK} becomes 
$\delta_V H = -2\pi \int_{\mathcal{H}^+} V^+ T_{++} + 2\pi \int_{\mathcal{H}^-} V^- T_{--},$where $\mathcal{H}^{\pm}$ are respectively the future and past Rindler horizons corresponding to the half-space. This fact, together with the monotonicity of relative entropy, was used in \cite{Faulkner:2016mzt} to prove the averaged null energy condition in general relativistic quantum field theories on Minkowski spacetime.} To see this in more detail, let us consider a general integral of the form
\beq \label{int1}
I_{\pm; \mu_1,\cdots\mu_m} = \lim_{B\to 0}\int_{-\infty}^{\infty} \frac{ds}{4\sinh^2(\frac{s+i\epsilon}{2})} \oint_0^{2\pi} B \,d\theta\,e^{\pm i\theta} \rho_R^{-is/2\pi}\,\cO_{\mu_1\cdots\mu_m}(B,\theta,y)\,\rho_R^{is/2\pi}.
\eeq
We now wish to perform the $\theta$-integral, which looks like a daunting task in a general reference state $\psi$. However,  note that operator $\cO$ is approaching the entangling surface in the $B\to 0$ limit, and we can therefore use the following important fact: in an infinitesimal Euclidean neighborhood of the entangling surface (much smaller than the scale associated with the extrinsic curvature of the entangling surface, or other scales associated with $\psi$), we can treat Euclidean modular evolution as being local, even for a non-trivial state and subregion! See \cite{Faulkner:2017vdd, Balakrishnan:2017bjg} for discussion about this\footnote{A heuristic argument for this is that in the $B\to 0$ limit, we can zoom-in to an infinitesimal neighborhood of the entangling surface (much smaller than the scale of extrinsic curvature of the entangling surface, and away from the other sources and operator insertions in the path-integral). In this region, $\xi = \pa_{\theta}$ is a symmetry of the Euclidean path-integral. We thank Tom Faulkner for multiple discussions about applying the methods of \cite{Faulkner:2016mzt} to non-local modular hamiltonians. }. In other words, in the $B \to 0$ limit we can approximately re-write the operator $\cO_{\mu_1,\cdots, \mu_m}(B,\theta, y)$ as:
\beq
\cO_{\mu_1 \cdots, \mu_m}(B,\theta,y) \simeq e^{-iw\theta} \rho_R^{-\theta/2\pi} \cO_{\mu_1 \cdots, \mu_m}
%{\nu_1\cdots \nu_m}
( B,0, y) \rho_R^{+\theta/2\pi}, 
\eeq
where $w$ the the modular weight of $\cO$, which can be defined in terms of holomorphic and anti-holomorphic coordinates $x^{\pm} = r e^{\pm i \theta}$ near the entangling surface, as the number of $+$ indices minus the number of $-$ indices on $\cO$. By shifting the $s$ contour by $s \to s+i\theta$, we can rewrite equation \eqref{int1} as
\beq \label{int2}
I_{\pm,w} = \lim_{B\to 0}\int_{-\infty}^{\infty}ds\,B \oint_0^{2\pi} d\theta\, \frac{1}{4\sinh^2(\frac{s+i\theta}{2})} e^{i(-w\pm 1)\theta} \rho_R^{-is/2\pi}\,\cO_{\mu_1\cdots\mu_m}(B,0,y)\,\rho_R^{is/2\pi}.
\eeq
We can now perform the $\theta$ integration, by using
\beq
\oint_0^{2\pi} d\theta\, \frac{1}{4\sinh^2(\frac{s+i\theta}{2})} e^{\pm im\theta} =  2\pi m e^{\mp ms} \Theta(\pm s) + \cdots,
\eeq
where the $\cdots$ denotes potential contact terms which have delta function support at $s=0$, and will not be relevant presently, as these contributions vanish in the $B\to 0$ limit. Now let us specialize to a two-index tensor $\cO_{\mu_1\mu_2}$. To see the potential enhancement, consider, for instance, $I_{-,w=-2}$ which corresponds to $\cO_{--}$:
\beq \label{int3}
I_{-,w=-2} = 2\pi \lim_{B\to 0}\int_{0}^{\infty}ds\,B  e^{-s} \rho_R^{-is/2\pi}\,\cO_{--}(B,0,y)\,\rho_R^{is/2\pi}.
\eeq
Using locality of modular flow for $Be^s  \ll \ell_K$ (where $\ell_K$ is the length-scale associated with the extrinsic curvature of the entangling surface), the modular flow on $\cO$ gives a Jacobian factor of $e^{2s}$, resulting in an overall $Be^s$ inside the integral, which becomes $O(1)$ when $s \sim \ln B$. Hence, the $s$-integral gives an enhancement in this limit. Of  course, we have no control over modular flow beyond this region of integration, and therefore we cannot do the integral explicitly -- but at the very least, the above enhancement is guaranteed to give a finite contribution from the region $ s \sim \ln B,\;Be^s \ll\ell_K$. We can similarly argue that $I_{+,w=+2}$ (corresponding to $\cO_{++}$) also gets enhanced. On the other hand, we find that all the other components of $\cO$ (such as $\cO_{+-},\;\cO_{\pm i},\;\cO_{ij}$) do not show enhancement in the accessible region of integration, and we expect that they do not receive enhanced contributions from the non-local region $B e^s \gg \ell_K$ either.\footnote{This follows from the usual expectation that correlators decay at large modular flow, see \cite{Faulkner:2017vdd} for example. } At any rate, for now we will leave equation \eqref{dXK} as it is, without performing the $\theta$-integral etc. -- the purpose of the above discussion was only to convince the reader that the operator $\delta_VH_{R}$ is in general non-vanishing in the $B\to 0$ limit. However, the above enhancement arguments will be crucial in section \ref{sec3}, where will apply them to bulk modular flow 
%near the HRRT extremal surface 
in the holographic setup.

A couple of further comments are in order. Firstly, note that the authors of \cite{Faulkner:2016mzt} included the unitary in the definition of the state while we consider it to be part of the definition of the modular hamiltonian. With this definition $\delta_V \psi=0$, because the state doesn't change under a surface translation and we get that the entropy variation is just:
\begin{equation}
\delta \delta_V S(\rho_{R})=\langle \delta_V H_{R} \rangle_{\delta \psi} \label{dSdK}.
\end{equation}
%\ack{What about the mod. Ham term, or equivalently the contact term? } \AL{Think about it better but I think it is OK} 
Secondly, we emphasize that while the shape deformation of the entangling surface is implemented by a Euclidean deformation of the metric, equation \eqref{dXK} ultimately yields a Lorentzian operator -- the Euclidean angular dependence, $\theta$, should be interpreted in the Heisenberg picture, and so $T(B,\theta,y)$ represents a non-local operator on $R$, at $x^0 = 0$. 
%should In this equation even if it naively looks like the stress tensor has support in both $R,\bar{R}$, we are actually thinking of the angular evolution  [THIS MIGHT BE EXPLAINED BETTER IN AN APPENDIX WHERE THIS IS DERIVED?]. 

\subsection*{Bulk perspective}
We now turn to shape deformations of the entangling surface for a holographic CFT with an Einstein gravity dual. In the bulk, this quantity is simple to compute -- using the Ryu-Takayanagi formula and its covariant generalization, it is given by the second variation of the area of the extremal surface  :
\begin{equation}
\delta \delta_V S(\rho_{R})=\frac{1}{4 G_N}\delta \delta_V A_{ext}, \label{RT}
\end{equation}
where as before, $\delta_V$ represents the shape deformation of the boundary entangling surface, and  now $\delta$ represents the change in the bulk geometry as a result of the state deformation in the CFT. In computing the right hand side, it is important that we account for the fact that the bulk extremal surface changes under a shape deformation, because we are simultaneously also considering a metric deformation . Let $g$ be the asymptotically AdS bulk metric dual to the state $\psi$, and let $\cS$ be the original bulk extremal surface corresponding to the boundary region $R$. Further, let $v^{I}(y^i)$ be the bulk vector field (a priori defined on $\cS$) which parametrizes the deformation of the bulk extremal surface under a shape deformation $V$ of the boundary subregion. The vector field $v^I$ satisfies the extremality condition\footnote{We are using Gaussian normal coordinates $(x^I, y^i)$ adapted to the original extremal surface $\cS$ in the bulk. Here, $y^i$ ($i=1, \cdots, d-1$) are coordinates along $\cS$ and $x^I$ ($I=1,2$) label directions orthogonal to it. In these coordinates, the original extremal surface $\cS$ sits at $x^I=0$. Additional details can be found in Appendix A.} (see Appendix A for further details):
\beq\label{ec}
\delta_{V} K_I=-\eta_{IJ} \gamma^{ij}\nabla_i\nabla_j v^J +  \gamma^{ij} R_{i(I; L)j } v^L - K_{I;ij}K_{L}^{ij}v^L= 0,
\eeq
on the original extremal surface at $\cS$, and approaches $V^I$ at the asymptotic boundary and $K^I_{ij}$ is the extrinsic curvature of ${\cal S}$. It is in principle possible to solve the above differential equation on $\cS$ subject to the asymptotic boundary condition to obtain $v^I$ in terms of $V^I$. 

Returning to equation \eqref{RT}, the change in the area of the extremal surface under the deformation $v^I$ is captured by the extrinsic curvature (up to boundary terms which are not important for our discussion):
\begin{equation} \label{d1A}
\delta_V A=\int_{\cS} d^{d-1}y \sqrt{\gamma}\; K_I  v^I,~~~ K^I \equiv \gamma^{i j} {\cal L}_{v^I} \gamma_{i j}.
\end{equation}
In this way, the second variation $\delta \delta_V A_{ext}$ of the area of the bulk extremal surface is given by\footnote{More explicitly, we have $$\delta \delta_V A_{ext}= \int_{\cS} d^{d-1}y\sqrt{\gamma}\, \left(  \frac{1}{2}\gamma^{ij}v^I\nabla_I\delta g_{ij} + \gamma^{ij}\delta g_{Ij} \nabla_{i} v^{I}\right), $$which is equivalent to equation \eqref{dA} up to unimportant boundary terms.}(see Appendix A for further details):
\beq \label{dA}
\delta \delta_V A_{ext}= \int_{\cS} d^{d-1}y \sqrt{\gamma} \delta K_I v^I, 
\eeq
where $v^I$ satisfies equation \eqref{ec}. 
%\ack{I restored the explicit expression in the footnote because I find the notation $\delta K_I$ somewhat abstract, and it's helpful to stare at the explicit formula to make sense of it.}

%\AL{This might need some change, do we have to talk about JLMS? I guess I wanted to talk more about EE because it seems more natural to consider and might speak to a broader audience.} \ack{I agrere.}

So far we have only considered the classical Ryu-Takayanagi entropy, but in what follows we will also be considering quantum deformations. In this case, the quantum corrections to Ryu-Takayanagi should be included:
\begin{equation}
\delta \delta_V S(\rho_{\psi,R})=\frac{\delta \delta_V A}{4 G_N}+\delta \delta_V S(\rho_{bulk,r} ), \label{QRT}
\end{equation}
where $r$ is the bulk subregion enclosed between the boundary subregion $R$ and the extremal surface $\cS$, and $S(\rho_{bulk,r} ) $ is the von Neumann entropy of the bulk quantum fields in $r$. Using arguments similar to those given around \eqref{dSdK}, we therefore arrive at
\begin{equation}
\langle \delta_V H_{R} \rangle_{\delta \psi} =\frac{\delta \delta_V A}{4 G_N}+ \langle \delta_V H_{bulk,r} \rangle_{\delta \psi_{bulk}}, \label{JLMS}
\end{equation}
where $\delta \psi_{bulk}$ denotes the change in the quantum state of the bulk fields resulting from the boundary state deformation $\delta \psi$. Of course, this is equivalent to the JLMS formula \cite{Jafferis:2015del} relating the boundary modular Hamiltonian to the sum of the bulk modular Hamiltonian and the area operator\footnote{The  statement $\text{tr} \delta H \rho=0$ is still true for the JLMS hamiltonian: $H_{R,\psi}=A_{\psi} \hat{I}+\int \sqrt{\gamma} \gamma^{a b} (\hat{\gamma}_{a b} -\hat{I}\gamma_{a b}^{\psi})$. However, note that this operator is state independent, it stays the same under a shift in the background: $\delta H_{R,\psi}=\delta A_{\psi}-\int \sqrt{\gamma} \gamma^{a b} \delta \gamma_{ab}^{\psi}=0$ , so $\langle \delta H_{\psi} \rangle_{\rho}=\text{tr} \delta \rho =0$.}:
\begin{equation}
H_{R}=\frac{\hat{A}}{4 G_N}+H_{bulk,r}  \label{JLMS0}.
\end{equation}
This formula is valid to order $G_N^0$ and should be thought as being evaluated in states $\psi+\delta \psi$ which are small deformations around a background state.  At this point it might be worth noting what we mean by $\delta \psi$. In this setting, we have in mind deformations of the state which are semi-classical -- these deformations can involve turning on a boundary source directly for the stress tensor\footnote{For simplicity we want to keep the boundary metric at $x^0=0$ fixed.} or other fields or just inserting some particles, as long as the overall change in the boundary stress tensor is small compared with $\frac{1}{G_N}$. If we focus on classical gravity, we want $\delta \langle T_{bdy} \rangle \sim \lambda G_{N}^{-1}, \lambda \ll 1$, but we might also consider quantum corrections, whose energy is $O(1)$. Given any semi-classical change in the state, we will have a corresponding classical change in the metric $\delta g$.  In \cite{Jafferis:2015del}, it was also argued from \eqref{JLMS0} that the boundary modular flow in $R$ and the bulk modular flow in $r$ are equivalent, a fact which will play a crucial role in section \ref{sec3}.

The goal of this paper is to prove equations (\ref{QRT}), \eqref{JLMS} from equation (\ref{dXK}), without using the replica trick. The reason why this will be possible is that in (\ref{dXK}), the stress tensor is integrated on a codimension-$1$ surface (instead of on the whole spacetime), which leads to a simple holographic dual for this operator. To understand how this works,  we need one last ingredient -- a gravitational identity from the Hollands-Iyer-Wald formalism.

\subsection{Hollands-Iyer-Wald formalism}

In order to make further progress in understanding the bulk-dual of (\ref{dXK}), we will need to recall the Hollands-Iyer-Wald (HIW) formalism \cite{Iyer:1995kg,Hollands:2012sf} (see also \cite{Lashkari:2015hha}), which can be used to relate ``gravitational'' quantities in the boundary CFT (i.e., involving the boundary stress tensor) with bulk quantities. Let $\bs{\epsilon}_{m_1\cdots m_n}$ be the $(d+1-n)$-form
\beq
\bs{\epsilon}_{m_1\cdots m_n} = \frac{1}{(d+1-k)!} \sqrt{g}\;\epsilon_{m_1\cdots m_n m_{n+1}\cdots m_{d+1}} dx^{m_{n+1}}\wedge  \cdots \wedge dx^{m_{d+1}},
\eeq
where we will use boldface notation for differential forms. For our purposes, the most important aspect of the HIW formalism is that the symplectic form $\bs{\omega}$ of the bulk gravitational theory satisfies the following purely gravitational identity  :
\beq \label{WF0}
\int_{\Sigma} \bs{\omega}(\delta g, \mathcal{L}_{X}g )  =   \int_{\pa\Sigma}\bs{\chi}(\delta g,X)- \int_{\Sigma} \bs{E}(g,\delta g,X),
\eeq
%\begin{equation}
%\int_{\partial \Sigma} n^{\mu} V^{\nu} \delta \langle T_{\mu \nu} \rangle=\int_{\Sigma} (n^{a} v^{b} E_{a b}(\delta g)+n_c V^c \delta g^{a b} E_{a b})+\int_{\Sigma} w(\delta g,{\cal L}_v g)
%\end{equation}
%where $w$ is the canonical symplectic form of the gravitational field (we will explore it in more detail soon), $E_{a b}$ is the Einstein tensor (with the cosmological constant included) and $v$ is some (in principle arbitrary) extension of the boundary diffeo to the bulk \AL{[I think the second term is OK]}. .
%\AL{write it as $\delta C$}
where $\Sigma$ is an arbitrary codimension-1 Cauchy surface in the bulk, and $X$ is an arbitrary vector field. Further, $\bs{\chi}$ is a $(d-1)$ form (to be defined below), and we have
\beq
 \bs{E}(g,\delta g,X)= -X^c \bs{\epsilon}_c \delta g^{ab} E_{ab}(g) + 2X^a\delta \left( E_{ab}(g) \bs{\epsilon}^b\right),
 \eeq
%where %$E^{(1)}_{mn}(\delta g)$ is the Einstein tensor including the cosmological constant term, linearized around the background metric $g$:
%\beq
%E^{(1)}_{mn}(\delta g) = \frac{1}{16 \pi G_N } \left(\delta R_{mn} - \frac{1}{2} \delta \left( R g_{mn}\right) + \Lambda \delta g_{mn} \right),
%\eeq
where $E_{ab}(g)$ is proportional to the non-linear equation of motion (including the cosmological term) for the bulk metric $g$. Equation \eqref{WF0} expresses gravitational quantities on $\Sigma$ in terms of the gravitational ``charge" on $\pa \Sigma$ and is the analog of Gauss' law for gravity. At no point have we used the equations of motion here. The bulk surface $\Sigma$ is in principle arbitrary. Finally, note that since this whole equation is linear in $\delta g$, we can write it as an operator equation by changing $\delta g \to \widehat{\delta}g$.

In this paper, we will set up formalism which works for a holographic CFT with a general gravity dual, but for concreteness we will focus on the case of Einstein gravity in the bulk. In this case, we can give explicit formulas for the various quantities appearing in equation \eqref{WF0}. The covariant gravitational symplectic 2-form density $\bs{\omega}$ in Einstein gravity is given by \footnote{This is equivalent, up to unimportant boundary terms, to the \emph{canonical} symplectic form \cite{BurnettWald}, which is essentially the gravitational version of the usual symplectic form $w= (\delta_1 p ~ \delta_2 x- \delta_2 p ~\delta_1 x)$ in classical mechanics. }
\beq
 \bs\omega(\delta_1g, \delta_2g) = \frac{1}{16\pi G_N} \bs{\varepsilon}_a P^{a(bc)d(ef)}[g] \left(\delta_1g_{bc} \nabla_d \delta_2g_{ef}-\delta_2g_{bc} \nabla_d \delta_1g_{ef}\right),
\eeq
where $P$ is the following tensor built out of the background metric:
\beq
P^{abcdef}[g]= g^{dc}g^{ea}g^{fb} - \frac{1}{2}g^{da}g^{eb}g^{fc} - \frac{1}{2}g^{dc}g^{ef}g^{ab}-\frac{1}{2}g^{df}g^{ea}g^{bc} + \frac{1}{2}g^{da}g^{ef}g^{bc}.
\eeq
The $(d-1)$-form $\bs{\chi}$ given by
\beqn
\bs{\chi}(\delta g, X)   &=& \delta \bs{Q}_X - i_X\bs{\Theta}(\delta g)\\
&=& \frac{1}{16\pi G_N} \bs{\varepsilon}_{ab}\left(\delta g^{ac}\nabla_{c}X^b - \frac{1}{2}{\delta g^c}_c\nabla^aX^b+ X^c\nabla^b{\delta g^a}_c - X^b\nabla_c\delta g^{ac} + X^b\nabla^a{\delta g^c}_c\right), \nonumber
\eeqn
where, as before, $X$ is an arbitrary vector field. Finally, we have 
\beq
E_{ab} = \frac{1}{16\pi G_N} \left(R_{ab} - \frac{1}{2}R g_{ab} + \Lambda g_{ab} \right). 
\eeq

%\AL{commented out one sentence, about the need to assume background extremality}
In the next section, we will see that by picking $\Sigma$ to be a suitable cylindrical tube which 
%surrounds the bulk extremal surface and 
ends on $\pa R_{\bbdy}$ at the asymptotic boundary, we can ``integrate in'' equation \eqref{dXK} into the bulk, thereby constructing a bulk-dual for $\delta_VH_R$.

\section{Integrating in the modular hamiltonian}\label{sec3}
In this section, we will present our main calculation. As explained previously, we consider a general (not necessarily ball-shaped) subregion $R$ in a general state $\psi$ of a holographic CFT dual to a smooth AAdS bulk geometry $g$. We will now prove equation \eqref{JLMS}:
\beq \label{jlms1}
\langle \delta_V H \rangle_{\delta \psi}= \frac{1}{4G_N}\delta_V A_{ext} + \langle \delta_V H_{bulk,r}\rangle_{\delta \psi_{bulk}},
\eeq
where as before, $\delta_V$ denotes the shape deformation and $\delta \psi$ is a state deformation (such that the backreaction in the bulk is small). In proving equation \eqref{jlms1}, we are going to assume: (i) the extrapolate dictionary near the asymptotic boundary, and (ii) that the boundary modular flow of $R$ for the reference state $\psi$ is equivalent to the bulk modular flow in the bulk region $r$ for $\psi_{bulk}$.% \ack{Actually, I think we should say this at a later state, and not here.} and that this also holds for small deformations of the subregion.

We emphasize that (ii) is only an assumption about the \emph{background} modular flow and in particular is weaker than the RT formula, since we do not need to mention the area: the modular flow is generated by the \emph{full} modular hamiltonian, in terms of which the assumption reads $\fH_{\partial R}=\fH_{\partial r, bulk}$. For the special case of a ball-shaped region in the vacuum, this assumption is the usual matching of bulk and boundary symmetries \cite{Casini:2011kv}, and for more general states and subregions was shown to follow from RT in \cite{Jafferis:2015del}. However, our approach here is to take this equality of the background bulk and boundary modular flows as a starting point, and use it to prove equation \eqref{jlms1}. To highlight how weak this assumption is, this is a property that can be checked completely in the realm of bulk perturbation theory: it is a statement about an equivalence in the code subspace, while the area contribution is not something calculable just from the code subspace, see \cite{Harlow:2016vwg} for some discussion about this. In other words, it is a property similar to bulk locality, in the sense which is present as long as there is a holographic dual, independently of the corresponding gravitational theory. 
%\AL{Omitted the next sentence, since it is a little confusing}
%Actually, our derivation below will reproduce equation \eqref{jlms1} up to a term which involves the linearized Einstein equation for $\delta g$ around the background geometry $g$ (including the bulk stress tensor term) and a term proportional to the extremality condition, which vanish on-shell. 

\subsection{Calculation}
\begin{figure}[t]
\centering
\includegraphics[height=6.5cm]{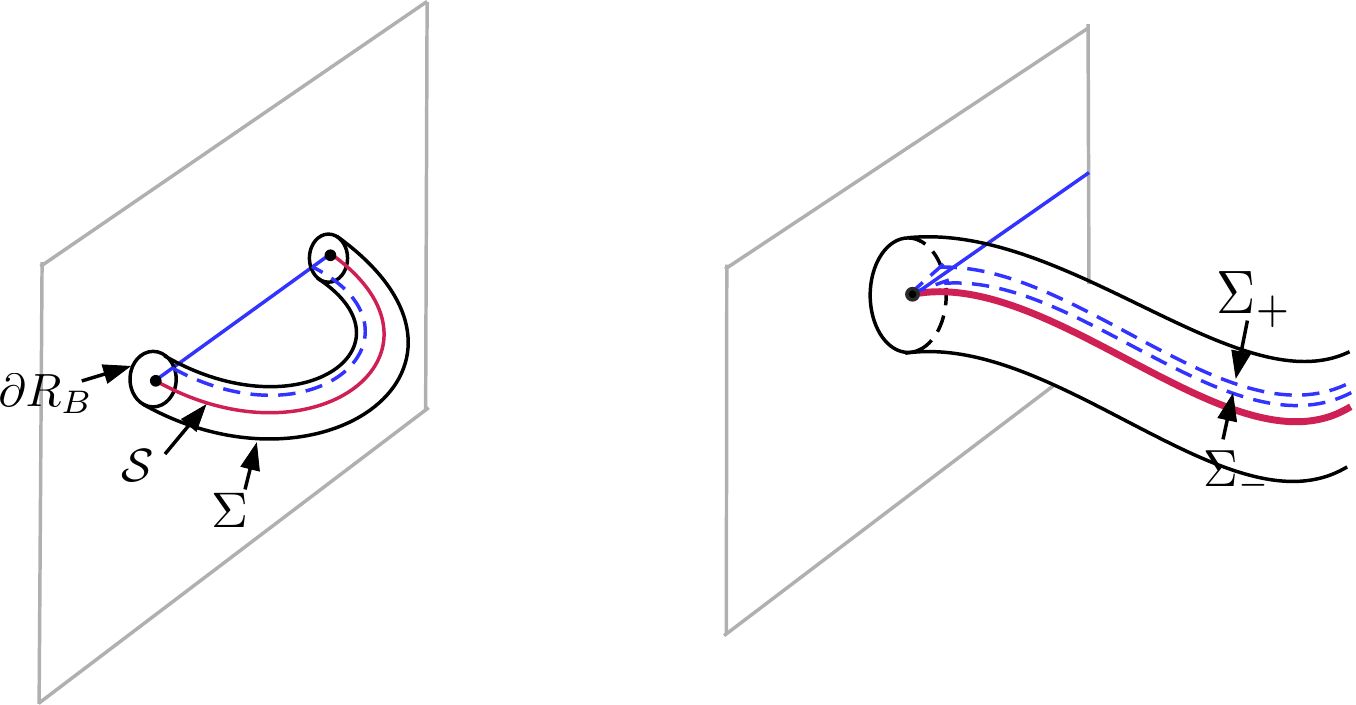}
\caption{\small{In the left panel is an illustration of the setup for constructing the bulk-dual of $\delta_VH_R$. The cylindrical tube $\Sigma$ surrounds the extremal surface $\cS$ (red curve) in the bulk. The solid blue line is the subregion $R$ in the boundary. The dashed blue line is the cut along $\tau = 0$ on $\Sigma$; its upper and lower boundary is $\Sigma_{+}$ and $\Sigma_-$ respectively as shown in the magnified picture in right panel.}\label{fig2}}
\end{figure} 
%For any finite $s$, this implies that, as $b \rightarrow 0$, $T_{\mu \nu}(s) \rightarrow T_{\mu \nu}(s=0)$ and for this to be reproduced by the previous equation, with the bulk modular and $H_{bulk}$, we need $\Sigma_b$ to be a tube of size $b$ surrounding $\partial r$.} 
Our starting point will be equation \eqref{dXK} in the boundary CFT. In order to proceed, we wish to rewrite equation \eqref{dXK} in terms of the bulk gravitational variables using equation \eqref{WF0}:
\beq \label{WF}
\int_{\Sigma} \bs{\omega}(\hat{\delta} g, \mathcal{L}_{v}g )  =  \int_{\pa \Sigma} \bs{\chi}(\hat{\delta} g,v)-\int_{\Sigma} \bs{E}(g,\hat{\delta} g,v),
\eeq
where we have used linearity in $\delta g$ to write this as an operator equation. For most of the calculation below we can be general and take $\Sigma$ to be a cylindrical tube surrounding some bulk surface $\cS$, and ending on $\pa R_B$. Consistency with the boundary expression requires ${\cal S}$ to be the codimension $2$ surface invariant under modular flow and ${\cal S}$
%We will only require that $\cS$ is invariant under boundary modular flow, and 
satisfies the condition that full boundary modular flow w.r.t. $\partial R$ is equal to full bulk modular flow w.r.t. $\cS$ (as stated previously); then $v$ will be the deformation of $\cS$ under the boundary shape deformation $V$ (or more more precisely, some smooth extension of this vector field around $\cS$).
%$v$ is then some vector field defined in a neighborhood of $\cS$. 
Looking ahead, $\cS$ will be the undeformed bulk extremal surface, but at this moment we don't need to impose this condition. 
%Actually,  \ack{Commented out a previous comment here, which I found to be confusing. Restore as a footnote if req.}
%Because we are assuming the equality between bulk and boundary modular flows, since the boundary expression goes to zero when $s \ll \log B$ and this should be also true in the bulk,  $\cS$ corresponds to the surface which is invariant under modular evolution. \ack{Say that we have in mind the extremal surface.} 
We can describe this tube locally in Gaussian normal coordinates adapted to $\cS$ (see Appendix A); in these coordinates the metric takes the form:
\begin{equation}
g=d\rho^2+\rho^2 d\tau^2+\gamma_{i j}(y) dy^i dy^j+\cdots
\end{equation}
where the surface $\cS$ is located at $\rho = 0$, $\tau$ is the angular coordinate around $\cS$, and we have neglected to write the $O(\rho)$ corrections for simplicity, but they are important for all calculations and can be found explicitly written out in Appendix A. In these coordinates, the cylindrical tube $\Sigma$ will be taken to be the surface $\rho= b$, with a cut along $\tau = 0$ (i.e., $\varepsilon \leq \tau \leq 2\pi - \varepsilon$). Let us first focus on the boundary term on the right hand side of equation \eqref{WF}. The boundary $\pa \Sigma$ consists of three components: the piece at the asymptotic boundary will be called $\Sigma_0$,\footnote{In Fefferman-Graham coordinates, we can put a cutoff in the radial direction and pick the asymptotic boundary to be at $z=z_0$. Then, we want the asymptotic component of $\pa \Sigma$ to match with $\pa R_B$, which requires $ B  \sim z_0 b$.} while the two boundaries along $\cal{S}$ at $\tau = \varepsilon$ and $\tau = 2\pi - \varepsilon$ will be denoted by $\Sigma_+$ and $\Sigma_-$ respectively (see figure \ref{fig2}).  A simple calculation shows that from $\Sigma_0$, we get
\beq \label{WF2}
\int_{\Sigma_0} \bs{\chi}(\hat{\delta} g,v) = \oint_{\Sigma_0} V^{\mu} n^{\nu} :T_{\mu\nu}:(B,\theta, y), 
\eeq
and so we may rewrite equation \eqref{WF} as
\beq \label{WFnew}
\oint_{\Sigma_0} V^{\mu} n^{\nu} :T_{\mu\nu}:(b,\theta, y)  = \int_{\Sigma} \bs{\omega}(\hat{\delta} g, \mathcal{L}_{v}g )- \int_{\Sigma_+\cup \Sigma_-} \bs{\chi}(\hat{\delta} g,v)+ \int_{\Sigma} \bs{E}(g,\hat{\delta} g,v).
\eeq
Therefore, using equations \eqref{dXK} and \eqref{WFnew}, and picking $\Sigma_0$ to coincide with $\pa R_B$, we can re-write $\delta_VH_R$ as:
\beqn \label{dvK}
\delta_V H_{R} &=& \lim_{b\to 0} \int_{-\infty}^{\infty} \frac{ds}{4\sinh^2(\frac{s+i\epsilon}{2})}  \rho_{R}^{-is/2\pi}\,\Big\{-\int_{ \Sigma_{+}\cup \Sigma_-} \bs{\chi}(\hat{\delta} g,v)+\int_{\Sigma} v^mT^{bulk}_{mn} \bs{\varepsilon}^n \nonumber\\
& &\hspace{2.5cm}+\int_{\Sigma} \bs{\omega}(\hat{\delta} g, \mathcal{L}_{v}g )+ \int_{\Sigma} \widetilde{\bs{E}}\Big\} \rho_{R}^{is/2\pi},
\eeqn
where we have defined $\widetilde{\bs{E}} =\left( \bs{E} - v^mT^{bulk}_{mn}\bs{\epsilon}^n\right).$ We have almost managed to rewrite equation \eqref{dXK} in terms of bulk gravitational variables, except for the CFT modular flow in the above equation. At this point, we assume that in the background, CFT and bulk modular flows are equivalent, i.e., the action of the CFT modular flow in equation \eqref{dvK} on bulk operators is equivalent to bulk modular flow. With this replacement, it is immediately clear that the second term on the right hand side is equal to $\delta_V H_{bulk,r}$. We therefore rewrite equation \eqref{dvK} as
\beq\label{dvK2}
\delta_V H_R= \delta_V H_{bulk, r} + \lim_{b\to 0} \int_{-\infty}^{\infty} \frac{ds}{4\sinh^2(\frac{s+i\epsilon}{2})}  \rho_r^{-is/2\pi}\,\Big\{-\int_{ \Sigma_+ \cup \Sigma_-} \bs{\chi}(\hat{\delta} g,v)+ \int_{\Sigma} \bs{\omega}(\hat{\delta} g, \mathcal{L}_{v}g )+\int_{\Sigma} \widetilde{\bs{E}}\Big\} \rho_r^{is/2\pi}.
\eeq
Now let us focus on the remaining terms individually. Firstly, the last term on the right hand side above 
\beq
\mathcal{E}_{flow} \equiv   \lim_{b\to 0} \int_{-\infty}^{\infty} \frac{ds}{4\sinh^2(\frac{s+i\epsilon}{2})}\int_{\Sigma}  \rho_r^{-is/2\pi}\; \widetilde{\bs{E}} \;\rho_r^{is/2\pi},
\eeq
is proportional to the equations of motion for the background metric $g$ and the linearized equations for $\delta g$ (including the bulk stress tensor), and so clearly this term vanishes on-shell. However in the interest of generality, we will keep this term as it is for now. Next, let's consider the term with $\bs{\omega}$. Since this term is integrated over $\Sigma$, naively it might appear to vanish in the $b \to 0$ limit; in fact, the only terms inside $\bs{\omega}$ which survive in the $b\to 0$ limit are terms which get enhanced by modular flow, coming from the region $s\sim \pm \ln b$ in the $s$-integration. Following the discussion in section $2$, the enhancement will only be sufficient for terms which have $2$ (or more) $+$ or $-$ indices on $\hat\delta g$. Therefore, it suffices to only keep track of the terms proportional to $\hat{\delta} g_{++}, \nabla_i \hat{\delta} g_{++}, \nabla_+ \hat{\delta} g_{++}$ and $\hat\delta g_{--}, \nabla_i \hat{\delta} g_{--},\nabla_- \hat{\delta} g_{--}$; the remaining terms vanish in the $b\to 0$ limit. Note that we have assigned modular weights to covariant tensors, which transform nicely under coordinate transformations.\footnote{For example, there is a term proportional to $V^IK_{I;ij}\gamma^{ik}\gamma^{jl}\nabla_{k}\delta g_{l\pm}$ in $\bs{\omega}$, which when expanded out contains a term proportional to $\delta g_{\pm\pm}$. However, we do not include this term because the covariant object $\nabla_{k}\delta g_{l\pm}$ has weight $\pm 1$ under modular flow.} We will present the details of this calculation in Appendix B.  The result is:
\beqn\label{SF}
\int_{\Sigma}\bs{\omega}(\hat{ g}, \mathcal{L}_{v}g ) &=& \frac{1}{8\pi G_N}\int_{\Sigma} d^{d-1}y\,b \left(e^{-i\tau} \hat{\delta} g_{--} \delta_V (\sqrt{\gamma} K_+) + e^{i\tau} \hat{\delta}{ g}_{++} \delta_V (\sqrt{\gamma} K_-) \right )+\cdots
%\left(- \frac{1}{2}\gamma^{ij}\nabla_i\nabla_jV^- +  \gamma^{ij} R_{i(+; L)j } V^L - K_{+;ij}K_{L}^{ij}V^L\right)\\
%&+&\frac{1}{8\pi}\int_{\Sigma} \delta g_{++}\left(- \frac{1}{2}\gamma^{ij}\nabla_i\nabla_jV^+ +  \gamma^{ij} R_{i(-; L)j } V^L - K_{-;ij}K_{L}^{ij}V^L\right)+\cdots
%\int_{\Sigma}\bs{\omega}(\delta g, \mathcal{L}_{V}g ) &=& \frac{1}{8\pi}\int_{\Sigma} \delta g_{--}\left(- \frac{1}{2}\gamma^{ij}\nabla_i\nabla_jV^- +  \gamma^{ij} R_{i(+; L)j } V^L - K_{+;ij}K_{L}^{ij}V^L\right)\\
%&+&\frac{1}{8\pi}\int_{\Sigma} \delta g_{++}\left(- \frac{1}{2}\gamma^{ij}\nabla_i\nabla_jV^+ +  \gamma^{ij} R_{i(-; L)j } V^L - K_{-;ij}K_{L}^{ij}V^L\right)+\cdots.\nonumber
\eeqn
where $\delta_V K_{\pm}$ can be written explicitly in terms of the original metric and $v^I$ as in \eqref{ec} %\beq
%\delta_{v} K_I \equiv -\eta_{IJ} \gamma^{ij}\nabla_i\nabla_j v^J +  \gamma^{ij} R_{i(I; L)j } v^L - K_{I;ij}K_{L}^{ij}v^L,
%\eeq
and $\cdots$ denote terms which do not get enhanced and hence drop out in the $b\to 0$ limit. Crucially, the terms proportional to $\delta g_{\pm\pm}$ in equation \eqref{SF} (which do survive as $b\to 0$) are proportional to the extremality condition for $v^{\pm}$! Therefore, imposing the extremality condition on $v$ (and ${\cal S}$) eliminates these terms. However, we will not impose extremality at this stage -- we will denote these two contributions collectively as $\delta_V {\cal S}_{flow}$.

\begin{figure}[t]
\centering
\includegraphics[height=2.3cm]{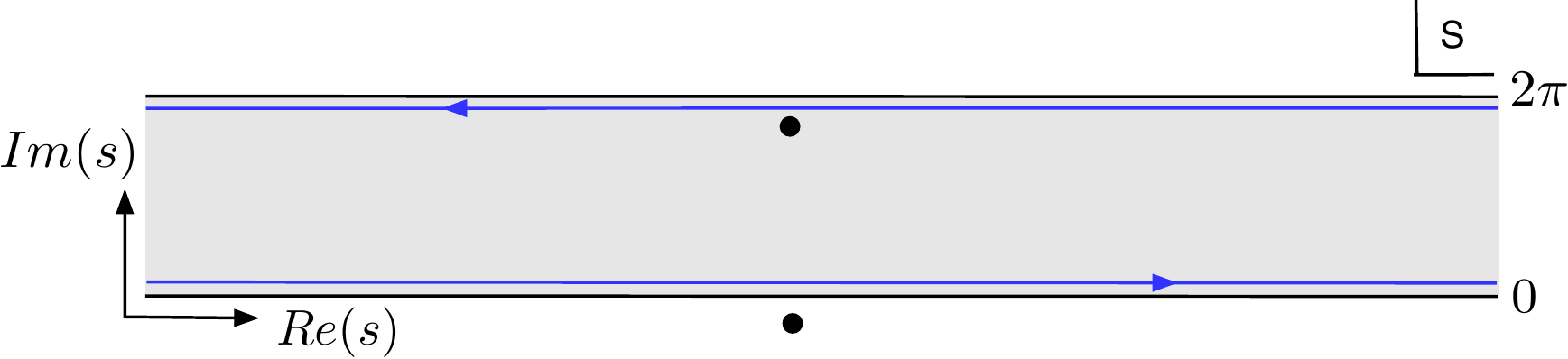}
\caption{\small{The strip $0 \leq Im(s) \leq 2\pi$ in the complex $s$-plane. The contour $C$ is shown in blue. The black dots are the poles of the kernel $\sinh^{-2}\left(\frac{s+i\epsilon}{2}\right)$. }\label{fig3}}
\end{figure}

Next, we move onto the $\bs{\chi}$ terms in equation \eqref{dvK2}. A short calculation shows (see Appendix B for details)
\beq
\int_{ \Sigma_{\pm}} \bs{\chi}(\hat{ g},v)=\frac{i}{8\pi G_N }\int_{\Sigma_{\pm}} d^{d-1}y\sqrt{\gamma} \left(\hat{\delta} K_- v^- -\hat{\delta} K_+ v^+ + \hat{\delta} g_{i j} \nabla^i v^j \right ) 
\eeq
where recall that $\langle \hat{\delta} K_{I}\rangle=\delta K_{I}$ was defined in equation \eqref{ec}. Now once again
% using the assumption of
using the locality of bulk modular flow near ${\cal S}$ (i.e., using the fact that $\tau=2\pi$ corresponds to $s=2\pi i$), we can rewrite the two terms (on $\Sigma_+$ and $\Sigma_-$) as
\beqn
\bs{\chi}-\mathrm{term} &=& -\int_{-\infty}^{\infty} \frac{ds}{4\sinh^2(\frac{s+i\varepsilon}{2})} \int_{ \Sigma_{+}\cup \Sigma_-}  \rho_r^{-is/2\pi}\bs{\chi}(\hat{\delta} g,v) \rho_r^{+is/2\pi} \\
&=&  \int_{C} \frac{ds}{4\sinh^2(\frac{s+i\varepsilon }{2})} \rho_r^{-is/2\pi} \int_{\Sigma_+}  \bs{\chi}(\hat{\delta} g,v) \rho_r^{+is/2\pi} \nonumber
%&=&  -\frac{i}{8\pi }\int_{C} \frac{ds}{4\sinh^2(s/2)} \rho_b^{-is/2\pi} \int_{\Sigma_+} d^{d-1}y\sqrt{\gamma} \left( \delta g^{i}_+\nabla_{i}V^+ + \frac{1}{2}V^+\nabla_+{\delta g^i}_i - (+ \to -)\right)\rho_b^{+is/2\pi} \nonumber
\eeqn
where $C$ is the following contour in the complex-$s$ plane (see figure \ref{fig3}):
\beq
C= \left(-\infty + i\varepsilon, \infty + i\varepsilon\right) \cup \left(\infty +i(2\pi -\varepsilon), -\infty +i(2\pi  -\varepsilon)\right).
\eeq
Using analyticity in the strip shown in figure \ref{fig3} together with the residue theorem, this contour integral picks up the double pole of $\sinh^{-2}(\frac{s+i\epsilon}{2})$ as $\varepsilon \to 0$, whose residue is equal to the commutator of $H_{bulk}$ with the integrand inside the $y$-integral:  
\begin{equation}
\bs{\chi}-\mathrm{term} = -\left[H_{bulk, r}, \bs{\chi}(\hat{\delta} g,v)\right]
\end{equation}  
Once again, the locality of the bulk modular flow near the RT surface implies that this commutator is determined by the local boost weight, which is $\pm 1$ for  $\hat{\delta} K_{\pm}$ \footnote{This can also be seen by expanding this object in covariant derivatives of $\hat{\delta} g$.} and $0$ for $\hat{\delta} { g}_{i j}$ , so we get:
%\beqn
%\bs{\chi}-\mathrm{term}&=& \frac{1}{4 } \int_{\Sigma_+} d^{d-1}y\sqrt{\gamma}\left(\delta g^{i}_I\nabla_{i}V^I + \frac{1}{2}V^I\nabla_I{\delta g^i}_i  \right)\nonumber\\
%&=& \frac{1}{4} \delta_V A_{ext.}
% \eeqn
 \beqn
 \bs{\chi}-\mathrm{term}&=&  \frac{1}{4 G_N}\int_{\Sigma_+} d^{d-1}y\sqrt{\gamma}\left(\hat{\delta} K_- v^- +\hat{\delta} K_+ v^+  \right)=\frac{1}{4 G_N} \delta_V \hat{A}[\cal S]+\cdots
 \eeqn
 where $\cdots$ denote terms localized on ${\cal S}$ which vanish when we take ${\cal S}$ to be extremal.

In conclusion, putting everything together, we find that
\beq \label{R1}
\delta_VH_R = \frac{1}{4G_N} \delta_V A[\cS] + \delta_VH_{bulk, r} +\cdots
%+\delta \cS_{flow} + \mathcal{E}_{flow},
\eeq
where the $\cdots$ correspond to terms that vanish when the equations of motion and the extremality conditions are satisfied and we have promoted this to an operator equation, because the state deformation was arbitrary, albeit with the standard caveat of small back-reaction. It is also useful to write down the corresponding equation for the full modular hamiltonian:
 \begin{equation}\label{R2}
\delta_V \fH_{\partial R}=\delta_V \fH_{\partial r,bulk}+\delta_V {\cal S}_{flow,r}-\delta_V {\cal S}_{flow,\bar{r}}+{\cal E}_{flow,r}-{\cal E}_{flow,\bar r}.
\end{equation}
We emphasize that so far, the only assumptions we have made are that for the original region $R$ and reference state $\psi$, the bulk and boundary modular flows are equivalent, together with the extrapolate dictionary. 

\subsection{Results}

Now we are ready to consider the various consequences of these formulas. 

\begin{itemize}
\item Let us begin by assuming that the bulk equations of motion are satisfied (by both $g$ and $\delta g$), $\cS$ is the undeformed extremal surface, and $v$ satisfies the extremality condition equation \eqref{ec}. In this case, the $\cdots$ in equation \eqref{R1} drop out, and we get
\beq\label{r3}
\delta_VH_R = \frac{1}{4G_N} \delta_V A_{ext} + \delta_VH_{bulk, r}, 
\eeq
which is the JLMS formula for the shape-deformed region. In particular, equation \eqref{r3} implies that if the bulk and boundary modular flows are equivalent
%JLMS formula is satisfied 
for some subregion $R$ (in some reference state $\psi$), and if the bulk geometry satisfies Einstein's equations, then a small shape deformation of the modular hamiltonian will satisfy JLMS. We want to highlight here the difference between JLMS and the equality of modular flows\footnote{Namely, that we did not assume JLMS in the background, but merely the weaker statement that the bulk and boundary modular flows are equal.}, because in the equality between modular flows there is no reference to the area term in the modular hamiltonian. In fact, from a quantum field theory perspective such a term localized on $\cS$ can be pretty subtle. However, in our result, it pops-out naturally in the bulk (from the minimal assumption of equality of background modular flows), and in particular is well-defined. Furthermore, equation \eqref{r3} implies that the equality between bulk and boundary modular flows continues to be true even for the deformed subregion. Importantly, since we assumed no special properties about the original subregion $R$ (for example, $R$ need not be ball-shaped in our arguments), we can now bootstrap this result to generate large shape deformations as well. This leads to an important corollary: if $\psi$ is the vacuum state, then \cite{Casini:2011kv} implies the equality of bulk and boundary modular flow for all ball-shaped regions; a ball-shaped region is therefore a natural candidate for the background subregion $R$. Since we can generate any compact subregion (with the topology of a ball) by deforming such a ball-shaped region, equation \eqref{r3} therefore implies the JLMS formula for such subregions of \emph{arbitrary shape} in the vacuum.

\item Alternatively, we could drop the assumption that $v$ satisfies the extremality condition \eqref{ec}, and instead assume that the boundary modular flow is equivalent to bulk modular flow \emph{even} in the deformed region $R+ \delta_VR$. This implies that the $\delta_V \mathcal{S}_{flow}$ terms in \eqref{R2} have to vanish -- this is because it is clear from the enhancement arguments given in section 2 that $\delta_V\mathcal{S}_{flow}$ would give a non-local operator with support in the bulk of $r$, so the equality between bulk and boundary modular flows necessarily implies 
\beq
\delta_V K_I=0,
\eeq
i.e., extremality for the perturbed surface $\cS + \delta_V\cS$. Since we can repeat these arguments perturbatively in the boundary shape deformations, the equality of modular flows does not allow for the corresponding deformations of the bulk surface $\cS$ to add extrinsic curvature with non-zero trace. We therefore expect that the extremality condition generally (for arbitrary subregions) follows from the equality between bulk and boundary modular flows! This is certainly true for any subregion with the topology of a ball in the vacuum, but for more general states we will leave it for future study.  

%   In this case, we can deduce the extremality condition $\delta_VK_I = 0$, because it is clear from the enhancement arguments given in section 2 that if the extremality condition were not satisfied, then $\delta_V\mathcal{S}_{flow}$ would give a non-local operator with support in all of $r$ -- this would contradict the assumption of equivalence between bulk and boundary modular flows in the deformed region.  

\item Next, we drop the assumption that the equations of motion are satisfied by the bulk metric deformation $\delta g$ (while $g$ still satisfies the background equation of motion), but we assume that the JLMS formula is satisfied for arbitrary regions; equivalently we may assume the Ryu-Takayanagi formula (including quantum corrections), which implies JLMS. In this case, we deduce from equation \eqref{R1} that we must have
\beq \label{eomr}
\mathcal{E}_{flow} \equiv  \lim_{b\to 0} \int_{-\infty}^{\infty} \frac{ds}{4\sinh^2(\frac{s+i\epsilon}{2})}\int_{\Sigma}  \rho_r^{-is/2\pi}\; \oint_{\Sigma} d\Sigma^m v^n E^{(1)}_{mn} \;\rho_r^{is/2\pi} = 0,
\eeq
where $E^{(1)}_{mn}$ is the linearized equation of motion, including the bulk stress tensor term. On the other hand, the enhancement from modular flow guarantees that the terms proportional to $E^{(1)}_{++}$ and $E^{(1)}_{--}$ in \eqref{eomr} are a priori non-trivial in the $b \to 0$ limit.\footnote{For instance in the case of local modular flow, we would have $\mathcal{E}_{flow} \sim- \int_{\mathcal{H}^+_{bulk}} v^+E^{(1)}_{++} + \int_{\mathcal{H}^-_{bulk}} v^-E^{(1)}_{--},$ where $\mathcal{H}^{\pm}_{bulk}$ are the bulk future and past horizons.} Given that the subregion $R$ is completely arbitrary, we expect that the only way equation \eqref{eomr} can be satisfied is if the null-null components of the linearized equation of motion are satisfied:
\beq
E_{\pm \pm}^{(1)} = 0, 
\eeq
although we have not attempted to prove this rigorously. If this can be shown, then this would prove that any AAdS geometry which satisfies the Ryu-Takayanagi formula (with quantum corrections) for first order state/metric deformations around the background geometry $g$, necessarily satisfies the linearized equations of motion $E_{\pm\pm}^{(1)}=0$ around $g$. Since the background geometry $g$ can be taken to be an arbitrary (not necessarily AdS) asymptotically AdS solution to the Einstein equation, this would then constitute a derivation of the full non-linear Einstein equation from entanglement in holographic conformal field theories! We end with the remark that this argument seems quite closely analogous to the original argument of Jacobson deriving the Einstein equation from the first law of thermodynamics \cite{Jacobson:1995ab}.

\end{itemize}

\section{Discussion} \label{sec4}
%\AL{Mark mention stating the assumptions more clearly, but I think we repeat them through the draft properly, theories with holographic duals and euqality of modular flows}
In this paper, we have combined the tecniques of \cite{Faulkner:2016mzt} with a purely gravitational identity from the Hollands-Iyer-Wald formalism to study the properties of entanglement entropy for subregions of arbitrary shape in conformal field theories with holographic duals. Starting from the equality between bulk and boundary modular flows in the background, we derived \eqref{R1} and \eqref{R2}, which upon the assumption of the equations of motion in the bulk, gave us the extremality condition and JLMS for arbitrary shapes. In the reverse direction, we were able to give an argument that any AAdS spacetime which satisfies the Ryu-Takayanagi formula must necessarily satisfy the non-linear Einstein equation. More precisely, we argued that the equation of motion integrated on an extremal surface should vanish for equation \eqref{eomr} to be true, but we have not given a detailed argument for the vanishing of the local equations of motion by ``inverting'' \eqref{eomr} -- we leave this to future work. Now, we would like to comment on some possible future directions or applications of this work. 
%\AL{Changed title}
\subsubsection*{It from modular flow?}
In the previous section, we considered three cases where we either \emph{assumed} the bulk equations of motion and derived JLMS and extremality, or vice versa. More generally, we expect the equality between modular flows in \eqref{R2} to impose $\delta_v {\cal S}_{flow}+{\cal E}_{flow}=0$. It is not clear to us if this equation has any other solution other than the two terms being individually zero. If there was a solution, it would be interesting to understand it better. If there is none, then both extremality and the equations of motion would follow form the equality of modular flows! 

In a related but slightly different direction, if we focus on the leading order in $G_N$ contribution around the RT surface (imposing equations of motion but not extremality for the shape deformation), it would seem that :
\begin{equation}
\delta_V \hat{H}_{R}=\frac{\delta_V \hat{A}_{ext}}{4 G_N}+\delta_V \hat{{\cal S}}_{flow}.
\end{equation}
This object is more bulk non-local than just the area operator and it might be possible to determine that $\delta_V \hat{{\cal S}}_{flow}$ is zero solely from comparing the properties of this object with the boundary modular hamiltonian. For instance, this operator doesn't seem to commute with operators which are space-like separated from the RT surface. However, it seems hard to make this statement more concrete.

At any rate, we seem to have a novel rewriting of the extremality condition in terms of a certain modular flow integral of the symplectic flux in the bulk (somewhat analogous to the discussion in \cite{Faulkner:2017tkh}), and it would be interesting to explore its physical interpretation further. Moreover, it is also of interest to give a more detailed derivation of the equations of motion based on the argument given in this paper. Finally, at various points in this paper we used the locality of modular evolution in an infinitesimal Euclidean neighborhood of the entangling surface, for modular times which are large-but-not-too-large; it would be useful to provide rigorous justifications. 

%\AL{Since this section is generally things to understand better mybae this goes well here? but change subtitle?}

\subsubsection*{Time dependent boundary time slices}
We introduced the shape variation through the Euclidean path integral, but we expect that in the boundary we can consider arbitrary Lorentzian Cauchy slices through analytic continuation. In these cases where the Cauchy slice is not a Euclidean section, we expect that this analytic continuation carries straightforwardly to the bulk: given a Lorentzian holographic mapping, we only need to continue the neightbourhood of the fixed point of bulk modular flow slightly into euclidean signature. We leave a more careful analysis of this for the future.

\subsubsection*{Higher derivatives}

Our discussion above was quite general, and except for the explicit computation of ${\bs\omega}$ and ${\bs \chi}$, we expect that it might in principle be straightforward to generalize to other theories of gravity. This might provide an alternative derivation of the formulas of \cite{Dong:2013qoa,Camps:2013zua} which does not rely on the replica trick (see also \cite{Haehl:2017sot} for progress along these lines). 
\subsubsection*{Quantum corrections}

Our discussion focused on the semiclassical regime, where gravitons can be thought of as free, spin-2 fields. That is, we have been working to order $O(G_N^0)$ in the entropy. Beyond that, we expect \cite{Engelhardt:2014gca,Dong:2017xht} that the position of the surface is shifted by a contribution proportional to the change in the bulk entanglement entropy. It would be nice to understand these corrections from our approach. We also expect some correction to the equality between modular flows \cite{Dong:2017xht} and our approach could provide useful to understand that better.

%\AL{Added}
\subsubsection*{Bulk vs boundary contributions in the $s$ integral }

We would like to highlight a feature of the calculation which might seem confusing at first\footnote{We thank Tom Faulkner for discussions about this.}. The boundary expression for $\delta_V H$ naively vanishes as $B \rightarrow 0$ but this suppression gets compensated by an enhancement at large modular time $s \sim \log B$. In the bulk, there is a similar contribution from $\omega,\delta_v H_{bulk}$, where large modular times give rise to finite terms. However, there is also the contribution from $\chi$ which is not suppresed in $b$, but, in the absence of modular flow, it would be zero because the contribution from $\Sigma_{\pm}$ would cancel. After modular flow, the non-trivial contribution of $\chi$ comes from the double pole (which occurs at small $s$). So, when going from the boundary to the bulk, one finds that there is some mixing on which modular times are contributing to the integral. This was already observed in \cite{Faulkner:2017tkh} and it might be related to the fact that the bulk solution might be ``more regular" than the boundary, in the spirit of \cite{Lewkowycz:2013nqa}. This seems to be related with the fact that the area operator is state independent, but we leave a more detailed analysis of this to the future. 

\subsubsection*{Modular flow versus replica trick}

Our approach had some similarities with the replica trick approach of \cite{Lewkowycz:2013nqa} -- there one assumed that the replica symmetry extended into the bulk, and here we assumed that the modular flow extends naturally into the bulk. Our surface ${\cal S}$ was defined in terms of the fixed point of this symmetry while in \cite{Lewkowycz:2013nqa} the RT surface is defined as the analytic continuation of the fixed point of replica symmetry. However, when doing the replica trick, one considers variations of the metric which look very singular, while our deformations have been rather mild, which makes it less constraining. 

Also, note that the our double-deformation of the entropy with respect to the state and shape morally resembles that of \cite{Dong:2017xht}, where they studied the dual of a different double deformation, i.e., a state variation together with a deformation of the Renyi parameter, which could also be integrated into the bulk. However, the integral in \cite{Dong:2017xht} was on a codimension-$0$ surface instead of codimension-$1$ because they had to integrate it in through the action. Nevertheless, it might be fruitful to better understand the connections between the two approaches.

%\subsubsection*{Possible generalizations}
%This idea of integrating in the integral of the stress tensor in the boundary tube could possible be generalize for other deformations of the state which correspond to conserved currents... \AL{Not clear} 
%It would be nice to understand  
%\AL{Not clear what do say about entanglement shadows}

%\AL{Not clear if there is anything to say about zero modes}
%\AL{A little bit confused about what to say about fluid dynamics, fluids normally don't have this sort of shape deformations, the zero modes are also there for the thermal state...}
%%%%%%%%%%%%%%%%%%%%%%%%%%%%%%%%%%%%%%%%%%%%
\section*{Acknowledgments}
We thank Joan Camps, Tom Faulkner, Antony Speranza and Mark Van Raamsdonk for useful discussions.   A.L. acknowledges support from the Simons Foundation through the It from Qubit collaboration. A.L. would also like to thank the Department of Physics and Astronomy at the University of Pennsylvania for hospitality during the development of this work. O.P's research supported by the Simons Foundation (\# 385592, Vijay Balasubramanian) through the It From Qubit Simons Collaboration, and the US Department of Energy contract \# FG02-05ER-41367.

\appendix

\section{Gaussian normal coordinates}
In the main text, we picked Gaussian normal coordinates adapted to the extremal surface for the \emph{background} metric; here we will list some further details about these coordinates. Note that we are \emph{not} fixing any particular gauge for the metric perturbations -- these will in general not preserve the form of the metric in Gaussian normal coordinates, but this choice of coordinates for the background simplifies the calculations. 

Given a codimension $2$ surface $\cal{S}$ with intrinsic coordinates $y^i$, we can denote the geodesic distance away from this surface by $\rho$. For small $\rho$, constant $\rho$ surfaces are tubes which we can parametrize by an angle $\tau$ and $(d-2)$ coordinates $y$. Our choice of the metric sets $g_{\rho \tau}=g_{\rho i} = 0$ everywhere on $\cS$. Furthermore, the fact that this is a codimension $2$ surface, forces $g_{\tau \tau}=\rho^2+O(\rho^4)$ and $g_{\tau i}=O(\rho^2)$. It will be convenient to work with null coordinates, $x^{\pm} = \rho e^{\pm i\tau}$.

 In these coordinates, the  metric takes the general form
\beqn
g &=& \delta_{IJ}dx^Idx^J + a_i(y)\varepsilon_{IJ}x^I\left(dx^Jdy^i +dy^idx^J\right)+\left(\gamma_{ij}(y) + 2 x^I K_{I;ij}(y)\right) dy^idy^j \\
&-&\frac{1}{3}R_{IK;JL}(y) x^Kx^Ldx^Idx^J + \frac{1}{3}R_{iK;LM}(y)\varepsilon_{IJ}\varepsilon^{LM}x^Ix^K \left(dx^Jdy^i +dy^idx^J\right) \nonumber\\
&+& \left(-4\delta_{IJ} a_ia_j + R_{i(I;J)j}(y) + K_{I;j}^{\ell}K_{J;i\ell}\right)x^Ix^Jdy^idy^j + O(x^3)
\eeqn
where $I, J \cdots= +, -$ denote indices perpendicular to $\cal{S}$ and and $i,j\cdots  = 1,2,\cdots, d-1$ denote indices along $\cS$. Further, $\delta_{IJ} = \left(\begin{matrix} 0 & 1/2\\ 1/2 & 0 \end{matrix}\right)$, and $\gamma_{ij}(y)$ is the metric in the directions parallel to $\cS$. The undeformed extremal surface is located at $x^I = 0$, with the extremality condition imposing $\gamma^{ij}K_{I; ij} = 0$. (The calculation can be carried out more generally without imposing this condition, but since we are ultimately interested in extremal background surfaces, we will take take $\gamma^{ij}K_{I; ij} = 0$ for simplicity).

Some useful Christoffel symbols evaluated at $x^{\pm} = 0$ are:
\begin{align}
\left. {\Gamma^m}_{IJ}\right|_{x=0} &= 0,\nonumber\\
\left. {\Gamma^I}_{iJ}\right|_{x=0} &= a_i\delta^{IK}\varepsilon_{JK},\nonumber\\
\left. {\Gamma^I}_{ij}\right|_{x=0} &= - \delta^{IJ}K_{J;ij}\nonumber\\
\left. {\Gamma^i}_{Ij}\right|_{x=0} &= \gamma^{ik}K_{I;kj}\nonumber\\
\left. {\Gamma^i}_{jk}\right|_{x=0} &= {\widehat{\Gamma}^i}_{jk} = \frac{1}{2}\gamma^{il}\left(\pa_i\gamma_{lj}+\pa_{j}\gamma_{li}-\pa_l\gamma_{ij}\right).
\end{align}
\section{Extremality Condition in Gaussian Normal Coordinates}
In this appendix, we derive the extremality condition and change in the area of the extremal surface (to first order in the both the shape and state deformation) in Gaussian normal coordinates for a general subregion in a general AAdS spacetime. 

 As explained in the main text, the shape deformation of the area and the extremality condition under a shape deformation are equivalent to:
\begin{equation}
\delta \delta_V A= \int dy \sqrt{\gamma} \left (\delta K^I+\frac{1}{2} \gamma^{i j} \delta \gamma_{i j}  K^I \right ) V^I ;~~~ \delta_V (\sqrt{\gamma}K^I)=0
\end{equation}
where if the background is of the previous form, the extrinsic curvature terms are given by:
\begin{equation}
\delta K_I=-K_I^{i j} \delta g_{i j}+\frac{1}{2} \gamma^{i j}\partial_I \delta g_{i j}- \gamma^{i j} \hat{\nabla}_{i} \delta g_{j I} = \frac{1}{2} \nabla_I \delta g^i_i-\nabla_i \delta g^{i}_I+ K_{i j} \delta g^{i j} \label{apdA}
\end{equation}
In particular, when $\delta g_{mn}=2\nabla_{(m} v_{n)}$ is a diffeomorphism, we can manipulate the covariant derivatives to write it in terms of the extrinsic curvatures, Riemann tensor and covariant derivatives of $v$ in the tangential direction.
%
%\begin{equation}
% \nabla_I \nabla_i V^i-g_{I J}\nabla_i \nabla^i V^J+\nabla_i \nabla_I  V^i+\# K_{i j} \nabla^i V^j
%\end{equation}

In this appendix, we will show this by explicit computation. In \cite{Mosk:2017vsz}, a similar expansion of the area was also considered.  
\subsubsection*{Extremality condition}
\newcommand{\cV}{v}
Let us first derive the extremality condition. In the interest of generality, we will begin by picking arbitrary coordinates $(x^I,y^i)$ for now, with the only requirement that the original extremal surface is located at $x^I=0$. Let the new extremal surface be located at $x^I = \cV^I(y)$. The induced metric is given by
\beq
h_{ij} = g_{IJ}(\cV,y) \pa_i\cV^I\pa_j\cV^J + 2g_{(i J}(\cV,y) \pa_{j)}\cV^J+ g_{ij}(\cV,y) .
\eeq
We can expand the induced metric for small $\cV$ as:
\beqn
h_{ij} &=& g_{IJ}(0,y) \pa_i\cV^I\pa_j\cV^J + 2g_{(i J}(0,y) \pa_{j)}\cV^J+ 2\pa_Kg_{(i J}(0,y) \cV^K\pa_{j)}\cV^J+ g_{ij}(0,y)\nonumber\\
&+& \pa_K g_{ij}(0,y)\cV^K+\frac{1}{2}\pa_K\pa_Lg_{ij}(0,y)\cV^K\cV^L+ \cdots
\eeqn
Note that we have dropped $O(\cV^3)$ terms above, because we are interested only in shape perturbations to linear order. We can make the following gauge choice here for convenience:
\beq
g_{iJ} (0,y) = 0.
\eeq
This is always possible; for instance, this is one of the conditions satisfied in the Gaussian normal coordinates we will use momentarily. Then the induced metric becomes
 \beqn
h_{ij} &=& g_{IJ}(0,y) \pa_i\cV^I\pa_j\cV^J + 2\pa_Kg_{(i J}(0,y) \cV^K\pa_{j)}\cV^J+ g_{ij}(0,y)\nonumber\\
&+& \pa_K g_{ij}(0,y)\cV^K+\frac{1}{2}\pa_K\pa_Lg_{ij}(0,y)\cV^K\cV^L+ \cdots
\eeqn
In order to ensure extremality, we need to vary with respect to $\cV$; the change in the induced metric is given by
\beqn
\delta_\cV h_{ij} &=& 2g_{IJ}(0,y) \pa_i\delta \cV^I\pa_j\cV^J + 2\pa_Kg_{(i J}(0,y) \delta \cV^K\pa_{j)}\cV^J+2\pa_Kg_{(i J}(0,y)  \cV^K\pa_{j)}\delta \cV^J\nonumber\\
&+& \pa_K g_{ij}(0,y)\delta \cV^K+\pa_K\pa_Lg_{ij}(0,y)\delta \cV^K\cV^L+ \cdots
\eeqn
In addition, we also need the relations
\beq
h^{ij} = g^{ij} - \cV^K(\pa_Kg_{i'j'})g^{ii'}g^{jj'}+\cdots
\eeq
\beq
\sqrt{ \mathrm{det}\,h_{ij}} =\sqrt{ \mathrm{det}\,g_{ij}}\left(1+ \frac{1}{2}g^{ij}\pa_Kg_{ij} \cV^K + \cdots \right)
\eeq
From here, we obtain
\beqn
\delta_V A &=& \int d^{d-1}y^i \sqrt{ \mathrm{det}\,g}\Big\{g^{ij}\left(2g_{IJ} \pa_i\delta \cV^I\pa_j\cV^J + 2\pa_Kg_{(i J} \delta \cV^K\pa_{j)}\cV^J\right.\nonumber\\
&+& \left. 2\pa_Kg_{(i J}  \cV^K\pa_{j)}\delta \cV^J+\pa_K g_{ij}\delta \cV^K+\pa_K\pa_Lg_{ij}\delta \cV^K\cV^L\right)\nonumber\\
&+& \frac{1}{2}(g^{ij}\pa_Kg_{ij} \cV^K)(g^{kl}\pa_Lg_{kl} \delta \cV^L) -g^{ik}g^{jl}\pa_Kg_{ij}  \,\pa_Lg_{kl} \cV^K\delta \cV^L)\Big\}.
\eeqn
Clearly, requiring that $\cV=0$ be an extremal surface requires $g^{ij}\pa_Kg_{ij} = 0$. This simplifies the above expression, and we obtain
\beqn
\delta_V A &=& \int d^{d-1}y^i \sqrt{ \mathrm{det}\,g}\Big\{ g^{ij}\left(2g_{IJ} \pa_i\delta \cV^I\pa_j\cV^J + 2\pa_Kg_{(i J} \delta \cV^K\pa_{j)}\cV^J\right.\nonumber\\
&+& \left. 2\pa_Kg_{(i J}  \cV^K\pa_{j)}\delta \cV^J+\pa_K\pa_Lg_{ij}\delta \cV^K\cV^L\right) -g^{ik}g^{jl}\pa_Kg_{ij}  \,\pa_Lg_{kl} \cV^K\delta \cV^L)\Big\}.
\eeqn

%We can now further simply by going to Gaussian normal coordinates adapted to the original extremal surface. In these coordinates, the bulk metric takes the general form
%\beqn
%g &=& \delta_{IJ}dx^Idx^J + a_i(y)\varepsilon_{IJ}x^I\left(dx^Jdy^i +dy^idx^J\right)+\left(\gamma_{ij}(y) + 2 x^I K_{I;ij}(y)\right) dy^idy^j \\
%&-&\frac{1}{3}R_{IK;JL} x^Kx^Ldx^Idx^J + \frac{1}{3}R_{iK;LM}\varepsilon_{IJ}\varepsilon^{LM}x^Ix^K \left(dx^Jdy^i +dy^idx^J\right) \nonumber\\
%&+& \left(-4\delta_{IJ} a_ia_j + R_{i(I;J)j} + K_{I;j}^{\ell}K_{J;i\ell}\right)x^Ix^Jdy^idy^j + O(x^3)
%\eeqn
%Some useful Christoffel symbols evaluated at $x^{\pm} = 0$ are:
%\begin{align}
%\left. {\Gamma^m}_{IJ}\right|_{x=0} &= 0,\nonumber\\
%\left. {\Gamma^I}_{iJ}\right|_{x=0} &= a_i\delta^{IK}\varepsilon_{JK},\nonumber\\
%\left. {\Gamma^I}_{ij}\right|_{x=0} &= - \delta^{IJ}K_{J;ij}\nonumber\\
%\left. {\Gamma^i}_{Ij}\right|_{x=0} &= \gamma^{ik}K_{I;kj}\nonumber\\
%\left. {\Gamma^i}_{jk}\right|_{x=0} &= {\widehat{\Gamma}^i}_{jk} = \frac{1}{2}\gamma^{il}\left(\pa_i\gamma_{lj}+\pa_{j}\gamma_{li}-\pa_l\gamma_{ij}\right).
%\end{align}
Now using Gaussian normal coordinates, we get the following extremality condition:
\beq
- 2\delta_{IJ} \widehat{\nabla}^2\cV^J + 2\gamma^{ij} (\pa_{I}g_{iJ})\pa_j\cV^J-2\gamma^{ij} \widehat{\nabla}_j\left(\pa_Kg_{iI}V^K\right) +\gamma^{ij}(\pa_I\pa_Lg_{ij}) \cV^L - \gamma^{ik}\gamma^{j\ell}(\pa_Kg_{ij} )(\pa_Ig_{k\ell} )\cV^K = 0.
\eeq
where $\widehat{\nabla}$ is the intrinsic covariant derivative in the original extremal surface. We can now covariantize the extremality condition by using 
\beqn
\pa_I\pa_J g_{ij} &=&  2\left( \delta^{KL}\varepsilon_{IK}\varepsilon_{JL} a_ia_j + R_{i(I;J)j} + K_{I;j}^{\ell}K_{J;i\ell}\right) + \cdots\nonumber\\
&=& 2\left( -4\delta_{IJ}a_ia_j + R_{i(I;J)j} + K_{I;j}^{\ell}K_{J;i\ell}\right) + \cdots
\eeqn
and we obtain
\beq
-\eta_{IJ} \gamma^{ij}\nabla_i\nabla_j\cV^J +  \gamma^{ij} R_{i(I; L)j } \cV^L - K_{I;ij}K_{L}^{ij}\cV^L= 0.
\eeq
This is the final form of the extremality condition we will work with. 

\subsubsection*{Area}
\newcommand{\dg}{\delta}
\newcommand{\dz}{\delta_V}

Next, we wish to compute $\delta_VA_{ext.}$, i.e. the change in the area of the extremal surface to linear order in the shape deformation and simultaneously linear order in the bulk metric deformation. The area of the RT surface is
\beq
A_{ext} = \int_{\cS} d^{d-1}y\,\sqrt{\mathrm{det}\,h_{ij}},\;\; h_{ij} = g_{mn}(x(y))\pa_{i} x^{m} \pa_{j}x^{n}
\eeq
If we deform the background geometry slightly, then this changes as 
\beq
\dg A_{ext} =  \frac12 \int_{\cS} d^{d-1}y\,\sqrt{\mathrm{det}\,h}\, h^{ij}\dg h_{ij}
\eeq
Importantly, $\dg h_{ij}$ has two terms:
\beq
\dg h_{ij} = \dg g_{mn} \pa_{i}x^{m}\pa_{j}x^{n} + \left( \delta x^p\pa_{p}g_{mn} \pa_{i}x^{m}\pa_{j}x^{n} +2g_{mn}\pa_{i} \dg x^{m}\pa_{j}x^{n}\right)
\eeq
where the first term is the change in the induced metric on the original surface, while the second term comes from the change in the minimal surface due to change in the bulk geometry. At the order we are working, we can discard the second term, because the original surface is extremal and so the change in the area coming from the second term should vanish. So we find
\beq
\dg A_{ext} =\frac12 \int_{\cS} d^{d-1}y\,\sqrt{\mathrm{det}\,h}\, h^{ij}\dg g_{mn} \pa_{i}x^{m}\pa_{j}x^{n}.
\eeq

We actually want to compute the first shape-derivative $\dz$-derivative of this term
\beqn
\dz \dg A_{ext} &=& \frac14 \int dy\sqrt{\mathrm{det}\,h}\, h^{ij}\dz h_{ij} h^{kl}\dg g_{mn} \pa_{k} x^{m} \pa_{l} x^{n}+ \frac12 \int dy\sqrt{\mathrm{det}\,h}\, \dz h^{kl}\dg g_{mn} \pa_{k} x^{m} \pa_{l} x^{n}\nonumber\\
&+&\frac12 \int dy\sqrt{\mathrm{det}\,h}\,  h^{kl}\dz\dg g_{mn} \pa_{k} x^{m} \pa_{l} x^{n}+ \int dy\sqrt{\mathrm{det}\,h}\, h^{kl}\dg g_{mn} \pa_{k} \dz x^{m} \pa_{l} x^{n}
\eeqn
where we have
\beq
\dz \dg g_{mn} = \cV^I\pa_{I}\dg g_{mn} 
\eeq
\beq\label{dh}
\dz h_{ij} =\cV^I \pa_{I}g_{mn} \pa_{i}x^{m}\pa_{j}x^{n} + 2 g_{In} \pa_{i}\cV^{I}\pa_{j}x^{n}=2 \cV^IK_{I;ij}
\eeq
So we obtain (now using Gaussian normal coordinates)
\beqn
\dz\dg A_{ext}&=& \frac12 \int dy\sqrt{\mathrm{det}\,\gamma}\, \cV^I \gamma^{ij}K_{I;ij} \gamma^{kl}\dg g_{kl} -  \int dy\sqrt{\mathrm{det}\,\gamma}\, \cV^IK_{I}^{ij}\dg g_{ij} \nonumber\\
&+& \frac{1}{2}\int dy\sqrt{\mathrm{det}\,\gamma}\,  \gamma^{ij}\cV^I\pa_I\dg g_{ij} + \int dy\sqrt{\mathrm{det}\,\gamma}\, \gamma^{ij}\dg g_{Ij} \pa_{i} \cV^{I} 
\eeqn
We can drop the first term because $\mathrm{Tr}\,K_I = 0$. Finally, covariantizing the remaining terms, we obtain
%\beq
%\dg\dz A=  \frac{1}{2}\int dy\sqrt{\mathrm{det}\,h}\,  h^{ij}V^I\pa_I\dg g_{ij} -  \int dy\sqrt{\mathrm{det}\,h}\, V^IK_{I}^{ij}\dg g_{ij}+ \int dy\sqrt{\mathrm{det}\,h}\, h^{ij}\dg g_{Ij} \pa_{i} V^{I} .
%\eeq
%For later use, we also rewrite the area in terms of covariant derivatives:
\beq
\dz \dg A=  \int_{\cS} d^{d-1}y\sqrt{\mathrm{det}\,\gamma}\left( \frac{1}{2} \gamma^{ij}\cV^I\nabla_I\dg g_{ij} +  \gamma^{ij}\dg g_{Ij} \nabla_{i} \cV^{I}\right) .
\eeq

It is easy to see that this is equivalent to (\ref{apdA}) up to integrations by parts in the $y$-directions. When integrating by parts, there is a boundary contribution which vanishes as long as the leading asymptotic of $\delta g_{i j}$ is fixed at the boundary (and thus the leading contribution comes from turning on the stress tensor). This boundary term doesn't vanish if we had $g_{I j}$ instead of $\delta g_{I j}$ and it is in fact the only contribution to the shape deformation of the entropy $\delta_V A$  (see for example equation (3.10) of \cite{Koeller:2015qmn}).
%We can think of this as $\delta \sqrt{|\gamma|} K^I V^I=\sqrt{|\gamma|} \delta \gamma K^I V^I+\sqrt{|\gamma}|} \delta K^I V^I$, with $\delta K^I_{i j}=\frac{1}{2} \partial_I \delta \gamma_{i j} - \hat{\nabla}_{(i} \delta g_{j) I}$, $\delta K=\gamma^{i j}\delta K_{i j}+\delta \gamma^{i j} K_{i j}$.  
\section{Details of the Symplectic 2-form and Boundary terms}
\newcommand{\varep}{\varepsilon}
In this appendix, we spell out the details of the $\bs{\omega}$ term and $\bs{\chi}$ term from equation \eqref{dvK2}.

\subsubsection*{$\bs{\omega}$-term}
Let us first consider the $\bs{\omega}$ term. Recall from section 2, that the gravitational symplectic form density is given by
\beq
\bs\omega(\delta_1g, \delta_2g) = \frac{1}{16\pi G_N} \bs{\ep}_a P^{a(bc)d(ef)}[g] \left(\delta_1g_{bc} \nabla_d \delta_2g_{ef}-\delta_2g_{bc} \nabla_d \delta_1g_{ef}\right),
\eeq
\beq
P^{abcdef} = g^{dc}g^{ea}g^{fb} - \frac{1}{2}g^{da}g^{eb}g^{fc} - \frac{1}{2}g^{dc}g^{ef}g^{ab}-\frac{1}{2}g^{df}g^{ea}g^{bc} + \frac{1}{2}g^{da}g^{ef}g^{bc}.
\eeq
We wish to integrate the symplectic form over the cylindrical tube $\Sigma$ (or radius $b$) surrounding the HRRT surface, with $\delta_2 g = \mathcal{L}_{V}g$. It is convenient to rewrite this integral as
\beq
\int_{\Sigma} \bs \omega(\delta_1g, \delta_2g) = \frac{1}{16\pi G_N} \int_{r=b} bd\theta d^{d-1}y\,\sqrt{\gamma}\; \left(\delta_1g_{bc} \delta_2\Pi^{bc}-\delta_2g_{bc} \delta_1\Pi^{bc} \right),
\eeq
where we have defined the \emph{covariant momentum}
\beq
\delta \Pi^{bc} = P^{r(bc)def} \nabla_d \delta g_{ef} = \frac{e^{-i\theta}}{2}P^{+bcdef} \nabla_d \delta g_{ef} +\frac{e^{+i\theta}}{2}P^{-bcdef} \nabla_d \delta g_{ef} .
\eeq
We are interested in the limit $b \to 0$. Let us evaluate the various components of the covariant momentum for small $b$:
\beq
\delta \Pi^{++} = 2 e^{-i \theta} \nabla_{-} \delta g_{--}- 2 e^{i \theta} \left(\nabla_{+} \delta g_{--} +\frac{1}{2} g^{ij} \nabla_{-} \delta g_{ij}\right)+O(b).
\eeq
\beq
\delta \Pi^{--} = 2 e^{+i \theta} \nabla_{+} \delta g_{++}- 2 e^{-i \theta} \left(\nabla_{-} \delta g_{++} +\frac{1}{2} g^{ij} \nabla_{+} \delta g_{ij}\right)+O(b).
\eeq
\beq
\delta \Pi^{ + - } =\frac{1}{2} e^{-i\theta} \left(g^{ij} \nabla_{-}\delta g_{ij} -2 \nabla_i \delta g^i_-\right) + \frac{1}{2}e^{+i\theta} \left(g^{ij} \nabla_{+}\delta g_{ij} -2\nabla_i\delta g^i_+\right)+O(b).
\eeq
\beqn\label{pipl}
\delta \Pi^{+i} &=&  e^{-i \theta} g^{ij}\nabla_{j} \delta g_{--} -  e^{i \theta}\left( g^{ij}\nabla_{+} \delta g_{-j}-g^{ij}\nabla_{-} \delta g_{+j}  +g^{ij}g^{kl} \nabla_{j} \delta g_{kl} \right)\nonumber\\
&+&g^{-i}\left(e^{-i\theta} \nabla_{-}\delta g_{--} - e^{+i\theta} \nabla_{+}\delta g_{--} \right) + O(b).
\eeqn
\beqn
\delta \Pi^{ij} &=& \frac{e^{-i\theta}}{2}\Big(2 g^{k(i}g^{j)l} \nabla_{l}\delta g_{-k} - g^{ik}g^{jl} \nabla_{-}\delta g_{kl}- 2 g^{ij} \nabla_{+}\delta g_{--} +2 g^{ij} \nabla_{-}\delta g_{+-}\nonumber\\
&-& g^{ij}g^{kl}\nabla_{k}\delta g_{l-}+g^{ij}g^{kl}\nabla_{-}\delta g_{kl}\Big)\nonumber\\
&+& \frac{e^{i\theta}}{2}\Big(2 g^{k(i}g^{j)l} \nabla_{l}\delta g_{+k} - g^{ik}g^{jl} \nabla_{+}\delta g_{kl}+2 g^{ij} \nabla_{+}\delta g_{-+} -2 g^{ij} \nabla_{-}\delta g_{++}\nonumber\\
&-& g^{ij}g^{kl}\nabla_{k}\delta g_{l+}+g^{ij}g^{kl}\nabla_{+}\delta g_{kl}\Big)+O(b)
\eeqn
Note that in the second line of equation \eqref{pipl}, we have kept an $O(b)$ term as it is relevant for our calculation (this term is $O(b)$ because $g^{-i} = 2 a^i x^-$), and neglected other $O(b)$ as well as higher order terms. 

Additionally, since one of the arguments of the symplectic 2-form is $\mathcal{L}_{V}g$, it is also convenient to  work out the various components of $\mathcal{L}_{V}g$ up to $O(b^2)$:
\beqn
\cL_{V}g_{IJ} &=& \delta_{IK} \pa_JV^K + \delta_{JK} \pa_IV^K+V^Kx^L\pa_K\pa_L g_{IJ} + O(b^2) \nonumber\\
&=& V^Kx^L\pa_K\pa_L g_{IJ} + O(b^2) .
\eeqn
\beqn
\cL_{V}g_{iI} &=& \delta_{IK}\pa_iV^K +a_i\varep_{KI}V^K + a_i \varep_{LK} x^L\pa_IV^K+ \pa_K\pa_Lg_{iI} x^LV^K + O(b^2)\nonumber\\
&=&\delta_{IK}\pa_iV^K +a_i\varep_{KI}V^K +  \pa_K\pa_Lg_{iI} x^LV^K + O(b^2).
\eeqn
\beq
\cL_{V}g_{ij} = 2K_{K;ij}V^K + \pa_K\pa_Lg_{ij} x^LV^K +a_i\varep_{LK}x^L \pa_jV^K +a_j\varep_{LK}x^L \pa_iV^K+  O(b^2).
\eeq

We are now in a position to extract the terms in the symplectic form which get sufficiently enhanced to survive in the $b \to 0 $ limit. These are terms which contain covariant tensors of modular weight $\pm 2$ (or higher), for e.g., $\delta g_{\pm\pm}$, $\nabla_i \delta g_{\pm \pm}$ and $\nabla_\pm \delta g_{\pm \pm}$ etc. For simplicity, let us focus on the -- terms.  We will use the notation $h^V = \mathcal{L}_Vg$ for convenience. We have two types of contributions in the symplectic flux density: (i) terms of the type $\delta g_{--} \Pi_{h^V}$ and (ii) terms of the type $h^V \Pi_{\delta g}$ (where we only keep objects with modular weight --2). The first type of contribution is given by:
\beqn
\delta g_{--} \Pi_V^-- &=&  \delta g_{--} \left\{ 2 e^{+i \theta} \nabla_{+} h^V_{++}- 2 e^{-i \theta} \left(\nabla_{-} h^V_{++} +\frac{1}{2} g^{ij} \nabla_{+} h^V_{ij}\right)\right\} \nonumber\\
&=& \delta g_{--} \left\{ 2 \left(e^{+i \theta}\pa_+h^V_{++}-  e^{-i \theta}\pa_-h^V_{++} \right)-g^{ij} \nabla_{+} h^V_{ij}\right\} \nonumber\\
&=&2\delta g_{--} \left(e^{+i \theta}\pa_+\pa_Kg_{++}-  e^{-i \theta}\pa_-\pa_Kg_{++} \right)V^K\nonumber\\
&-& \delta g_{--} \,g^{ij} \Big(V^I\pa_+\pa_Ig_{ij} -4K^l_{+,i}K_{I,jk}V^I +4a_{i} a_{j}V^-\Big).
\eeqn
where we have used 
\beqn
\nabla_+h^V_{ij} &=& \pa_+ h^V_{ij} - 2{\Gamma^k}_{+(i}h_{j)k} - 2{\Gamma^+}_{+(i}h_{j)+}\nonumber\\
&=&V^I\pa_+\pa_Ig_{ij} + 2a_{(i}\pa_{j)}V^--4K^l_{+,(i}K_{I,j)k}V^I - 4a_{(i}\left(\frac{1}{2}\pa_{j)}V^- - a_{j)}V^-\right). 
\eeqn
The second type of contribution is of the type $h^V \Pi_{\delta g} $:
\beqn
h^V \Pi_{\delta g} &=& h^V_{++} \left(2e^{-i\theta}\nabla_{-}\delta g_{--}  - 2e^{+i\theta} \nabla_+\delta g_{--}\right) + 2h^V_{i+}e^{-i\theta} g^{ij}\nabla_j\delta g_{--} - e^{-i\theta} g^{ij}h^V_{ij}\nabla_+\delta g_{--} \nonumber\\
&=& V^Kx^L \pa_K\pa_L g_{++} \left(2e^{-i\theta}\pa_{-}\delta g_{--}  - 2e^{+i\theta} \pa_+\delta g_{--}\right)-e^{-i\theta} g^{ij}h^V_{ij}\nabla_+\delta g_{--}\nonumber\\
&+&2\left(\frac{1}{2}\pa_iV^- - a_iV^-\right)e^{-i\theta} g^{ij}\left(\nabla_{j} \delta g_{--}- 2a_j\delta g_{--}\right). 
\eeqn
Note that we have also kept terms of the form $\nabla_+\delta g_{--}$ for completeness, and in the last line, the $a_j$ term comes from integrating by parts the second line of equation \eqref{pipl}. Combining all the terms together, we find that the $\delta g_{--}$ contribution (or more precisely the weight --2 contribution) to the symplectic form is given by:
\beqn
&=&\delta g_{--} \,\gamma^{ij} \Big(-V^I\pa_+\pa_Ig_{ij} +4K^l_{+,i}K_{I,jk}V^I -4a_{i} a_{j}V^- + \nabla_j\left(\pa_iV^- - 2a_iV^-\right)\Big) \nonumber\\
&=&\delta g_{--} \,\gamma^{ij} \Big(-V^I\pa_+\pa_Ig_{ij} +4K^l_{+,i}K_{I,jk}V^I  +\widehat{\nabla}_j\left(\pa_iV^- - 2a_iV^-\right)-2a^i\left(\pa_iV^-\right)\Big)
\eeqn
where note that we have integrated by parts along the $y^i$ direction, and it can be checked that the boundary term vanishes by the asymptotic boundary conditions. Finally, rewriting $\pa_+\pa_Ig_{ij}$ in Gaussian normal coordinates, we find that the coefficient of $\delta g_{--}$ is precisely the extremality condition.

\subsubsection*{$\bs{\omega}$-term using the canonical symplectic form}
Above, we used the \emph{covariant} version of the symplectic form. There is a second, quicker way to arrive at the above result using the \emph{canonical} version of the symplectic form, which we now briefly explain. In the canonical formalism, we have:
\begin{equation}
\bs{\omega}=\delta_1 h_{m n} \delta_2 p^{m n}-\delta_2 h_{m n} \delta_2 p^{m n}
\end{equation}
where $h,p$ are the induced metric and momenta in the tube $\Sigma$ surrounding the extremal surface $\cS$. The momentum is defined by:
\begin{equation}
p^{m n}=\sqrt{h} (K_{b}^{m n}-h^{m n} K_b)
\end{equation}
where the $b$ subindex denotes that this is the extrinsic curvature in the codimension $1$ surface. 

In this case, the rule for enhancement is that we only have to keep the $\delta h_{\pm \pm},\delta p_{\pm \pm}$ terms. Since $\delta_V h_{\pm \pm}=0$, i.e., the surface translation doesn't change the $\tau,\tau$ component of the metric, we only get a contribution from $\delta_V p^{\pm \pm}$. For simplicity, let us work with cylindrical coordinates, we have:
\begin{equation}
p^{\tau \tau}=\sqrt{\gamma} b^{-1} \gamma^{i j} K_{b,i j}^{\tau} \label{ptt}
\end{equation}

and thus its $v$ variation will be given by \footnote{In \eqref{ptt} we plugged the background value for the components of the metric which are not changed under $\delta_V$: $\delta_V g_{I J}=0$, so $\delta_V K_{i j}$ depends on both $\delta g_{i j},\delta g_{I i}$.}:
\begin{equation}
\delta_V p^{\tau \tau} b=  \delta_V (\sqrt{\gamma} K_{+}) e^{i \tau}+\delta_V(\sqrt{\gamma} \delta K_{-}) e^{-i \tau}
\end{equation}
where we used that $K^{\tau}_{b,ij}=K_{+,i j} e^{i \tau}+K_{-,i j} e^{-i \tau}$. In this way since $\delta g_{\tau \tau}=e^{2 i \tau} \delta g_{++}+e^{-2 i \tau} \delta g_{--}+2 \delta g_{+ -}$ , the only contribution which gets enhanced will be:
\begin{equation}
{ \bs \omega}=\delta g_{--} \delta_V (\sqrt{\gamma} K_{+}) e^{-i \tau}+\delta g_{++}\delta_V(\sqrt{\gamma} K_{-}) e^{i \tau} +...
\end{equation}

% even if $K_b^{\tau}$ has an extra contribution from $g^{\tau \tau} K_b^{\tau \tau}$ which is no a codimension $2$ surface extrinsic curvature, it doesn't contribute because its variation is zero, when contracted with $\delta g_{\tau \tau}$, the only terms that have the right $e^{i \tau}$ dependence are $\delta_V K_+ \delta g_{++}+\delta_V K_- \delta g_{--}$ .   
%
%\AL{This $\delta_V$ contribution is OK. There might be another term hiding that doesn't come from here and is proportional to $K$. This term is $\delta p_{i j} {\cal L}_v h^{i j}$ and basically the story is that ${\cal L}_v h^{i j}=K^{i j}$ and the $ p_{i j}=\sqrt{h} n_r s^{\tau \tau} \gamma_{i j} \Gamma^r_{\tau \tau}=\sqrt{\gamma} \gamma_{i j}$ in the background. There is a contribution to this term that goes like $\delta n_r$... which is not suppresed. In my newcalc.pdf, I mentioned how there is a term here that goes like $\delta g_{\tau \tau}$ but I had forgotten about the $\delta n_r$ that does get excited with a $\delta g_{zz}$...}. 

\subsubsection*{$\bs{\chi}$-term}
Now recall that
\beq \label{chiexp}
\bs{\chi}(\delta g, V)   = \frac{1}{16\pi G_N} \bs{\varepsilon}_{ab}\chi^{ab},\;\;\;\chi^{ab}= \left(\delta g^{ac}\nabla_{c}V^b - \frac{1}{2}{\delta g^c}_c\nabla^aV^b+ V^c\nabla^b{\delta g^a}_c - V^b\nabla_c\delta g^{ac} + V^b\nabla^a{\delta g^c}_c\right).
\eeq
Let us first focus on the contribution coming from the cut $\Sigma_{\pm}$. Here we get
\beq
\bs{\chi}(\delta g, V)   = \frac{1}{16\pi G_N}\sqrt{g} \,dy^1\wedge \cdots \wedge dy^{d-1}\times \frac{i}{2b} \left(\chi^{+-} - \chi^{-+}\right),
\eeq
where we have
\beqn\label{cpm}
\chi^{+-} &=& \left(\delta g^{+c}\nabla_{c}V^- + V^c\nabla^-{\delta g^+}_c - V^-\nabla_c\delta g^{+c} + V^-\nabla^+{\delta g^c}_c\right)\nonumber\\
&=&\left(\delta g^{+i}\nabla_{i}V^- + V^I\nabla^-{\delta g^+}_I - V^-\nabla_i\delta g^{+i}-V^-\nabla_I\delta g^{+I} + V^-\nabla^+{\delta g^i}_i+V^-\nabla^+{\delta g^I}_I\right),\nonumber
\eeqn
and similarly, 
\beqn\label{cmp}
\chi^{-+} &=& \left(\delta g^{-i}\nabla_{i}V^+ + V^I\nabla^+{\delta g^-}_I - V^+\nabla_i\delta g^{-i}- V^+\nabla_I\delta g^{-I} + V^+\nabla^-{\delta g^i}_i+V^+\nabla^-{\delta g^I}_I\right).\nonumber
\eeqn
Combining equations \eqref{cpm} and \eqref{cmp}, we find that the only terms which survive are given by
\beq
\bs{\chi}(\delta g, V)  =\frac{i}{8\pi G_N}\sqrt{\gamma}\,dy^1\wedge \cdots \wedge dy^{d-1} \left(\delta g^{i}_-\nabla_{i}V^-+\frac{1}{2}V^-\nabla_-{\delta g^i}_i - \delta g^{i}_+\nabla_{i}V^+ - \frac{1}{2}V^+\nabla_-{\delta g^i}_i+\delta g_{i j} \nabla^i v^j \right)
\eeq
which is the result used in the main text. Note that the $v^I K_I^{i j} \delta g_{i j}=\nabla^i v^j \delta g_{i j}$ term comes from the integration by parts of $\nabla_i \delta g^{I -}$ term, since we are integrating by parts in a codimension $2$ surface. 

Finally, the $\bs{\chi}$ term at the asymptotic boundary can be evaluated using Fefferman-Graham coordinates. We use the asymptotic expansions:
\beq
g = \frac{dz^2+ dx_{\mu}dx^{\mu}}{z^2} + \cdots ,
\eeq
\beq
v^I = V^I + \cdots, \;\;\;\hat\delta g_{\mu\nu}  = z^{d-2} \frac{16 \pi G_N}{d}  :T_{\mu\nu}: + \cdots.
\eeq
Substituting in equation \eqref{chiexp}, we find that in the $z\to 0$ limit the only contribution comes from the $v^I z\nabla_z \delta g^r_I$ term which gives 
\beq 
\int_{\Sigma_0} \bs{\chi}(\hat{\delta} g,v) = \oint_{\Sigma_0} V^{\mu} n^{\nu} :T_{\mu\nu}:(B,\theta, y).
\eeq
\providecommand{\href}[2]{#2}\begingroup\raggedright\endgroup

\end{document}